\documentclass[traditabstract,a4paper]{aa}
\usepackage{amsmath}
\usepackage{graphicx}
\usepackage{epsfig}
\usepackage{color}
\usepackage{txfonts}
\usepackage{natbib}
\usepackage[normalem]{ulem}
\usepackage[bookmarks=false,colorlinks=true, citecolor=blue, backref=false,breaklinks=true]{hyperref} 
\bibpunct{(}{)}{;}{a}{}{,} 

\newcommand*\diff{\mathop{}\mathrm{d}}

\begin{document}

\title{Are Cosmological Gas Accretion Streams Multiphase
and Turbulent?}

\authorrunning{Cornuault et al.}
\titlerunning{Are Gas Accretion Flows Turbulent?}

\author{Nicolas Cornuault\inst{1}\thanks{email: cornuault@iap.fr}
\and Matthew D. Lehnert\inst{1}
\and Fran\c{c}ois Boulanger\inst{2}\thanks{Research associate at the Institut d'Astrophysique de Paris}
\and Pierre Guillard\inst{1}}

\institute{Sorbonne Universit\'{e}s, UMPC Paris 6 et CNRS, UMR 7095,
Institut d'Astrophysique de Paris, 98 bis bd Arago, 75014 Paris, France
\and Institut d'Astrophysique Spatiale, CNRS, UMR8617, Universit\'{e}
Paris-Sud 11, B\^{a}timent 121, Orsay, France}

\abstract{
Simulations of cosmological filamentary accretion reveal flows (``streams'') of warm gas, T$\sim$10$^4$K,
which are efficient in bringing gas into
galaxies.  We present a phenomenological scenario where
gas in such flows -- if it is shocked as it enters the halo as we
assume -- become biphasic and, as a result, turbulent. 
We consider a collimated stream of warm gas that flows into a halo from an over dense 
filament of the cosmic web. The post-shock streaming gas expands because it has a higher
pressure than the ambient halo gas, and 
fragments as it cools. The fragmented stream forms a two phase medium:
a warm cloudy phase embedded in hot post-shock gas. 
We argue that the hot phase sustains the accretion shock. 
During fragmentation, a fraction of the initial kinetic energy of the infalling gas is converted into
turbulence among and within the warm clouds. The thermodynamic
evolution of the post-shock gas is largely determined by the relative
timescales of several processes. These competing timescales characterize
the cooling, the expansion of the post-shock gas, the amount of
turbulence in the clouds, and the dynamical time of the halo.
We expect the gas to become multiphase when the gas cooling and dynamical times are of the same order-of-magnitude.
In this framework, we show that this occurs in the important mass range of M$_\mathrm{halo}$$\sim$10$^{11}$ to 10$^{13}$ M$_{\sun}$, where the bulk of stars have formed in galaxies.
Due to expansion and turbulence, gas accreting along cosmic web filaments may eventually loose coherence and  mix with the ambient halo gas.  Through both the phase separation and ``disruption'' of the stream,
the accretion efficiency onto a galaxy in a halo dynamical time
is lowered. De-collimating flows make the direct interaction
between galaxy feedback and accretion streams more likely, thereby
further reducing the overall accretion efficiency. As we discuss, moderating the gas
accretion efficiency through these mechanisms may help to alleviate a
number of significant challenges in theoretical galaxy formation.
}


\maketitle

\section{Introduction}
\label{sec:intro}

The realization that we live in a dark matter dominated universe led to
the development of the first comprehensive theory of galaxy formation
\citep[e.g.][]{white78, fall80}.  These analytic models embedded
simple gas physics into the hierarchical growth of structure whereby
smaller halos merged over time forming successively more massive halos
\citep{white78}.  Despite the successes of this model in understanding
the scale of observed galaxy masses, it was soon realized that there
were a number of problems. The most significant is that modeled
galaxies form with a higher fraction of baryons than is observed
\citep[e.g.,][]{ferrara05, bouche06, anderson10, werk14}. This failure
was dubbed the ``over-cooling problem'' \citep{benson03}.

As numerical simulations allowed for galaxy growth to be coupled to
the development of large scale structure, they showed that much of
the accreting mass may penetrate into the halo as filaments of gas and
dark matter \citep{keres05, ocvirk08}.  Whether or not these streams
pass through a stable accretion shock as they penetrate the halo depends
on the mass and redshift of the halo \citep[][hereafter BD03 and DB06
respectively]{birnboim03, dekel06}. If the shock is not stable, the accretion flow
is ``cold''. 
This ``cold mode'' accretion in simulations occurs as streams of 
warm ($10^4$~K) gas entering the halo, smooth at kpc-scale and 
weakly coupled to the infalling dark matter filaments \citep{danovich15,
wetzel15}.  Cold mode accretion is efficient in reaching down to a few
tenths of a virial radius \citep{dekel06,
behroozi13}. The high efficiency of gas accretion in some simulations
leads to model galaxies with unrealistically high baryon fractions
emphasizing the over-cooling problem.  To alleviate the problem of
excess baryons in simulated galaxies, efficient outflows and feedback
were introduced \citep[e.g.,][]{hopkins12, hopkins16}. Feedback both
heats the gas in the halo, preventing it from cooling, and also ejects
gas from both the galaxy and halo lowering their total gas content.

The circum-galactic media of galaxies are certainly not devoid of gas,
perhaps containing up to approximately half of the total baryon content of
the halo \citep[e.g.,][]{werk14, peek15}. This gas is known empirically
to be multiphase. The multiphase nature of halo gas is most evident in
local high mass halos, those with masses on the scales of cluster or
groups. In clusters, for example, even at constant pressure, a very wide
range of gas phases are observed, from hot X-ray emitting gas to cold,
dense molecular gas \citep[e.g.,][]{jaffe05, edge10, salome11, tremblay12,
hamer16, emonts16}.  In galaxy halos, the detection of multiphase gas
is mostly through the absorption lines from warm neutral and ionized gas
\citep[$\la$10$^4$ to $\sim$10$^6$ K][]{} and dust via the reddening of
background galaxies and quasars \citep[][but see \citealt{pinto14} for the
detection of hot gas in X-ray emission lines]{menard10, peek15}. Outflows
from galaxies are also multiphase \citep[e.g.,][]{beirao15, heckman17}
and are likely crucial for creating and maintaining the multiphase gas in
halos \citep[e.g.,][]{gaspari12, sharma12, borthakur13, voit15, hayes16}.

There is only circumstantial evidence for smooth, collimated accretion
streams penetrating into galaxy halos \citep[e.g.,][ and
references therein]{martin15, bouche16, vernet17}. In analogy with
analyses of gas in halos and outflows, a phenomenological approach
may provide additional insights into the nature of flows and halos that
galaxy simulations are perhaps not yet achieving \citep[e.g.,][]{sharma10,
sharma12, sharma12a, singh15, voit15, voit15a, thompson16}.  The notion that over-cooling
remains a problem in simulations, that there is scant observational
evidence for the streams of the type currently simulated, and because high speed
collisions of gas can lead to multiphase turbulent media 
\citep{guillard09, guillard10, ogle10, peterson12, appleton13, alatalo15}, 
all motivated us to analyze gas accretion flows phenomenologically.

Just as with the explanation for the lack of cooling flows in clusters
\citep[e.g.,][]{peterson03, rafferty08}, our current understanding of
accretion flows in galaxy halos may also suffer from an overly simplistic
view of gas thermodynamics. In clusters, it is now understood that
heating and cooling are in approximate global balance, preventing
the gas from cooling catastrophically \citep[e.g.,][]{rafferty08,
sharma12, sharma12a, mcCourt12, zhuravleva14, voit15a}. In analogy
with gas in cool core clusters, and in contrast to what a number of
cosmological simulations currently show, the gas in streams may not
cool globally.  
Instead, if the gas in streams is inhomogeneous and subject to thermal 
and hydrodynamic instabilities \citep[e.g.,][]{sharma10}, the differences 
in cooling times between the gas phases will lead to fragmentation of the gas. 
If streams are unstable, their gas will not remain monophasic or laminar. 
Thus, our goal in this paper is to
investigate the question posed in the title: ``Are Cosmological Gas
Accretion Streams Multiphase and Turbulent?''. If yes, the gas energetics
may regulate the gas accretion efficiency.  \citet{dekel13} estimated the
penetration efficiency over large halo scales at z$\sim$2 of $\sim$50\%
but this estimate only considered macroscopic processes that may influence
the accretion efficiency. Because heating and cooling are controlled by the 
mass, energy and momentum exchanges between the gas phases, a careful investigation of
the gas physics on microscopic scales is required to investigate whether those microphysical  
processes may further reduce the efficiency of gas accretion onto galaxies. 

To investigate the question posed in the title, we begin by presenting a qualitative
sketch of our scenario, make a quantitative investigation of the impact of expansion in the post-shock gas on an accretion flow, and then discuss the case where the post-shock gas has small fluctuations in density and temperature, developing a criterion for when the gas will fragment (\S\,\ref{sec:motivation}). In \S\,\ref{sec:consequences}, we  discuss the consequences of the formation of a multiphase flow on the thermodynamic evolution of the stream
after it penetrates the halo. To gauge
the astrophysical pertinence of our model, we analyze the evolution of
an idealized gas accretion stream into a halo of 10$^{13}$\,M$_{\sun}$
at $z$=2 (\S\,\ref{sec:specificcase}). In Section~\ref{sec:discussion},
we discuss why simulations may be missing some ingredients necessary for
modeling accretion shocks robustly and outline a few simple consequences
of our proposed scenario.

\section{Our framework for multiphased streams}\label{sec:motivation}

The idea that gas in halos is multiphase has been suggested for decades
\citep[e.g.,][]{binney77, maller04}. More recent studies of halo gas
attribute the development of multiphase gas to the growth of local thermal
instabilities \citep[e.g.,][]{sharma10} or galaxy outflows
\citep[e.g.,][]{thompson16, hayes16}. Thermal instability is only
relevant when the cooling time of the unstable gas is of the same order-of- magnitude or smaller than
the dynamical time of the halo \citep[e.g.,][]{sharma12, mcCourt12}. In ambient halo
gas and outflows from galaxies, the gas must often meet this requirement
given that they are observed to be multiphase. However, in an accreting stream of gas, it is
difficult to understand how the gas might achieve such a balance in heating and cooling.
We may have to consider other processes to determine if it is possible for streams
themselves to become multiphased as they flow into the halo.
In the following sections, we examine the physics of gas flowing into halos from cosmic web filaments.

\begin{figure*}[t]
\resizebox{\hsize}{!}{\includegraphics{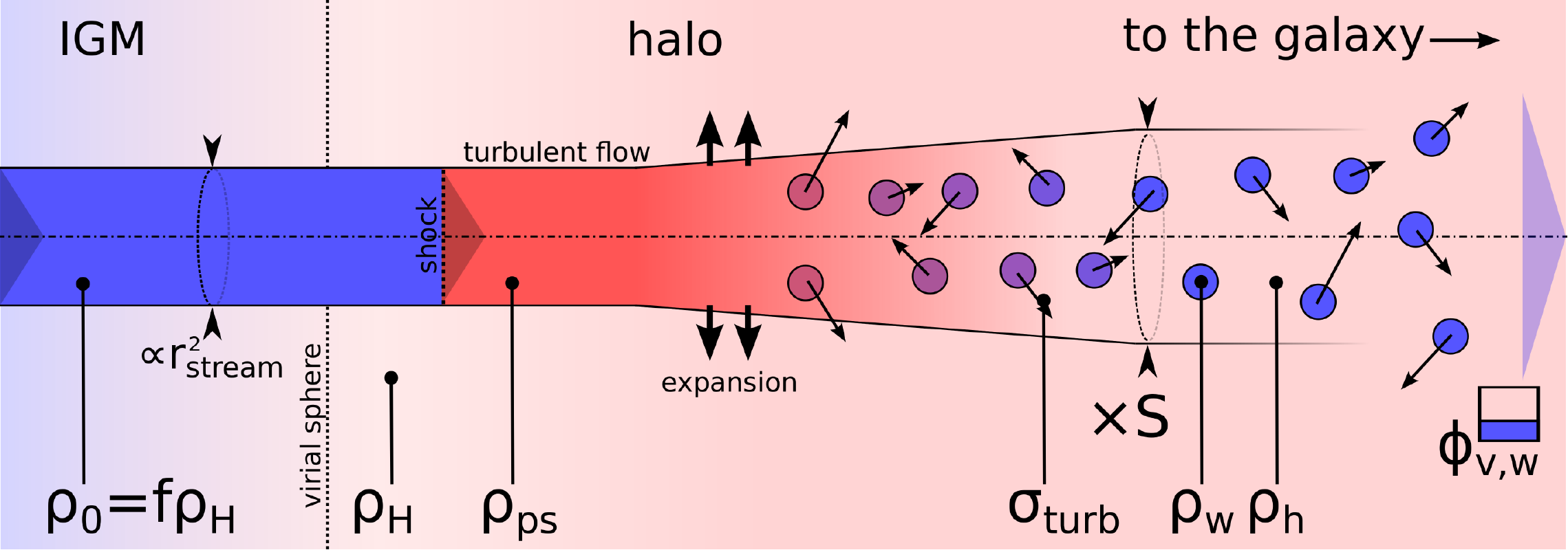}}
\caption{Sketch of our phenomenological picture of flows of gas passing
through a virial shock. The initial inflowing gas (in blue) is
over-dense relative to the halo gas at the boundary by a factor $f$
(=$\rho_0/\rho_\mathrm{H}$ or the stream density divided by the density
of the ambient halo gas at the virial radius), shocks at the boundary
between the inter-galactic medium (labeled as ``IGM'') and the hot halo
gas (labeled as ``halo'').  The persistent virial shock and the higher
pressure of the post-shock gas compared to the ambient halo gas, allows
the flow to expand after being shocked (labeled as ``expansion'').
The post-shock gas may become unstable, fragmenting to form
a biphasic medium. The fragmentation enables a fraction of the initial
momentum and energy of the stream to be captured as turbulent clouds of
warm gas with a dispersion, $\sigma_\mathrm{turb}$, and a volume-filling
factor, $\phi_\mathrm{v,w}$.  The clouds may move beyond the initial
radius of the stream, de-collimating the flow. Eventually, the hot post-shock gas mixes with
the ambient halo gas which prevents it from cooling further. See text and
the Appendix for definitions of the variables.}
\label{fig:sketch}
\end{figure*}

\subsection{Qualitative sketch of our specific framework}
\label{subsec:qualitativesketch}

We briefly qualitatively outline our scenario of gas accretion through streams, 
sketched in Fig.~\ref{fig:sketch}, introducing the concepts
developed later in the paper.
We consider a collimated stream of warm gas with a temperature of
10$^4$\,K that penetrates into a dark matter halo filled with hot
gas at the halo virial temperature. The smooth
density distribution of the hot gas follows the density distribution
of the underlying dark matter halo. The ambient halo gas has a long
cooling time and has constant density and temperature during the flow. The speed of the
flow is set to the virial velocity of the halo and is highly supersonic relative to
the sound speed of the gas within the stream. In the following we introduce the 
three basic physical ingredients of our modeling of streams.

As the stream penetrates at the virial radius, we assume it is shock-heated whenever the hot halo gas provides the necessary pressure support for sustaining the shock (\S\,\ref{subsec:expansion}).  The post-shock gas is over-pressurized relative to the ambient halo gas and expands, invalidating the classical one-dimensional analysis
of streams \citep[DB06,][]{mo10}. The hot post-shock gas mixes with the ambient halo gas,
which prevents much of the gas initially in the stream from cooling completely to
form a monophasic post-shock stream. We further posit that the post-shock gas will develop
inhomogeneities due to, for example, non-planarity or obliqueness of the shock-front
or through inhomogeneities in velocity and/or density of the stream before it is shocked. The fragmentation of the gas into hot and warm cloudy phases is central to our scenario. If certain physical conditions are met, the expanding inhomogeneous gas will cool and fragment, forming
a two phase medium -- a hot phase with an embedded warm cloudy phase
(\S\,\ref{subsec:expansion}). Density and velocity inhomogeneities in the post-shock
gas may be amplified through gas cooling, leading to the formation  of a multiphased
flow (see \S\,\ref{subsec:diffcooling}). 

We expect that part of the kinetic energy of the infalling gas will be
converted to turbulence within the warm clouds and random cloud-cloud motions
\citep[e.g.,][]{hennebelle99, kritsuk02, Heigl2017}. If the level of turbulence is
high or if the clouds are formed while the gas is expanding, the warm clouds may spread beyond the initial boundary of the
collimated inflowing stream (Fig.~\ref{fig:sketch} and see \S\,\ref{sec:consequences}).
If these conditions are met, the stream will de-collimate.
The thermodynamic evolution of
the post-shock gas in the stream is largely determined by the
timescales of several relevant processes -- gas cooling, expansion of the hot phase, 
dynamical time of the halo, and the stream disruption due to turbulent motions --
which we quantify in the remainder of this and the next section.

\begin{figure}[t]
\resizebox{\hsize}{!}{\includegraphics{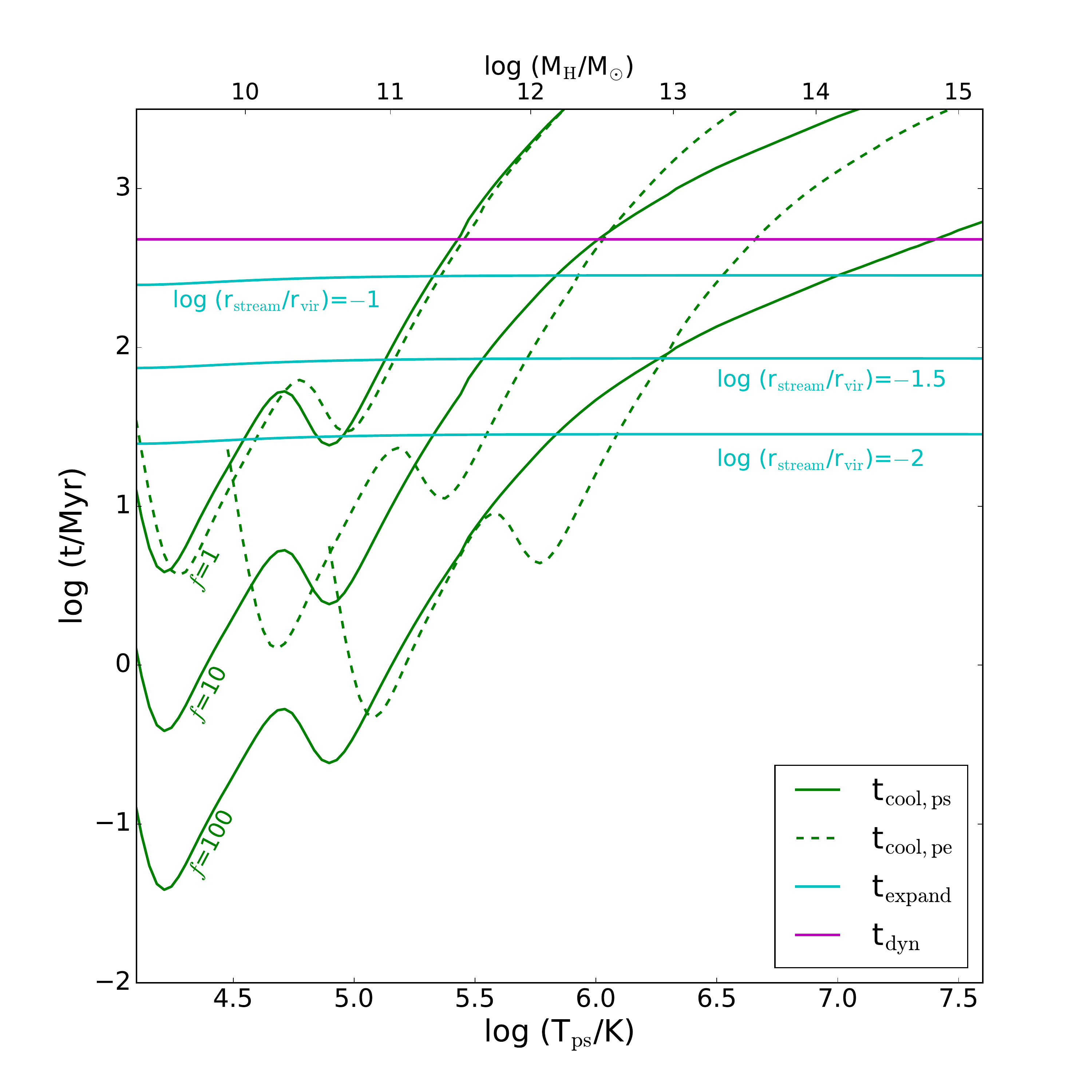}}
\caption{Timescale comparison of the competing relevant processes of cooling,
expansion, and halo dynamics, for different post-shock temperatures (i.e.,
different halo masses), at redshift $z=2$.  Cooling times (green)
are times of isobaric cooling from a given temperature-density couple.
The solid green curves indicate the instantaneous post-shock cooling time,
$t_\mathrm{cool,ps}$, and the green dashed curves indicate the post-expansion
cooling time, $t_\mathrm{cool,pe}$ -- the cooling time of the post-shock
after it has expanded sufficiently to reach pressure equilibrium with
the ambient halo gas.  The cooling time curves for the post-expansion gas are
truncated at the minimum temperature spanned by the curves, 10$^4$\,K. The cooling time curves were computed for three
values of the initial over-density $f$, as indicated by green labels on
the left side of the panel.  Expansion times (solid cyan lines)
are shown for three different relative filament radii, log$_\mathrm{10}$
r$_\mathrm{stream}$/r$_\mathrm{vir}$=$-$1, $-$1.5, and $-$2. 
The dynamical time, $t_{\rm dyn, halo}$, which is independent of
halo mass, is indicated by the horizontal magenta line.}
\label{fig:tscales_vs_MH}
\end{figure}

\subsection{The Impact of Expansion on the Phenomenology of Accretion Flows}
\label{subsec:expansion}

Since the pioneering work of \citet{white78} decades ago, the accretion
of gas onto galaxies or into halos has been analyzed in one, radial dimension,
assuming homogeneity of the flow \citep[see][]{mo10}. The one-dimensional
approximation is also used for streams, where it is valid only if
one can ignore lateral expansion. To know if this approximation is valid,
one has to compare the expansion to cooling time of the gas within the
flow. Historically, even the analysis of the stability of the accretion
shock neglected the possibility that the post-shock gas may expand
into the ambient halo gas gas \citep[DB06; ][]{mo10}. In our study,
we reconsider the sustainability of the accretion shock that occurs
as the infalling filament collides with the ambient halo gas. To conduct
this analysis, we compare the cooling time of the post-shock gas to
various other measures of its dynamical evolution. For simplicity, we
only compare the cooling time to the expansion time of the post-shock
gas into the surrounding ambient halo gas and the dynamical time of
the halo. Intuitively, one can understand that if the cooling time is
significantly shorter than either the expansion or dynamical times,
the stream after the shock will quickly cool down and maintain much of
its integrity. Such a shock can be unstable and this is essentially the
textbook situation that has been considered already \citep{mo10}. If
the cooling time of the post-shock gas is significantly longer than either the
expansion or dynamical times, then the post-shock gas will mix the ambient
halo gas further supplying it with hot gas and entropy. This is another
textbook example that has been analyzed one-dimensionally \citep[DB06;
][]{mo10}. However, if the cooling time is of the same order-of-magnitude
of the expansion and halo dynamical times,
a wider range of outcomes of the post-shock gas are
possible \citep[e.g.,][]{sharma10,sharma12a}. This is
the point of this paper -- to describe phenomenologically this more
complex regime.

We now consider the impact of expansion on the properties of the
post-shock gas. To estimate the cooling, expansion, and dynamical
timescales, we introduce two parameters specific to the flow, its
over-density relative to that of mean density of gas at the virial
radius, $f$, and the radius of the stream, $r_\mathrm{stream}$.
All variables used in this and subsequent sections are summarized in
Tables~\ref{tab:haloparams} and \ref{tab:modelvars} of the Appendix.
As we now show, the gas in the stream immediately after being heated
to a high post-shock temperature, $T_\mathrm{ps}$, has a pressure,
$P_\mathrm{ps}$, higher than the ambient halo gas, $P_\mathrm{H}$,
and thus expands into the surrounding ambient halo gas of density,
$\rho_\mathrm{H}$. All of the post-shock quantities are given by
the normal Rankine-Hugoniot shock relations (see Appendix for appropriate formulae).
The halo pressure at the virial radius is,

\begin{equation}
P_\mathrm{H}=\left(\frac{k_\mathrm{B}}{\mu m_\mathrm{p}}\right)\rho_\mathrm{H}
T_\mathrm{H}= \frac{(\gamma-1)}{2}\rho_\mathrm{H} v_\mathrm{vir}^2
=\frac{\gamma \mathcal{M}_1^2}{3f}P_0
\label{eqn:PH}
\end{equation}

\noindent
where $T_\mathrm{H}$ is the temperature of the ambient halo gas, which we 
have assumed to the be virial temperature. The
pressure ratio just after the shock is,

\begin{equation}
\frac{P_\mathrm{ps}}{P_\mathrm{H}}=\frac{2f}{\gamma \left(\gamma+1\right) \cdot g\left(\mathcal{M}_1\right)}
\end{equation}

\noindent where $g:M_1\mapsto (2\gamma/(\gamma-1)-M_1^{-2})^{-1}$. For
$\gamma=5/3$ and within the domain [1,$+\infty$], this function
rapidly decreases from $1/4$ to $1/5$. Hence the pressure ratio,
$P_\mathrm{ps}/P_\mathrm{H}$ is between 9/5 and 9/4 times the initial
stream over-density, $f$. Since $f$$\ga$1, the pressure of the post-shock
gas is higher than the pressure of the ambient halo. Thus, since the
pressure ratio is always greater than one, the flow will expand
into the ambient halo gas. 

\begin{figure}[t]
\resizebox{\hsize}{!}{\includegraphics{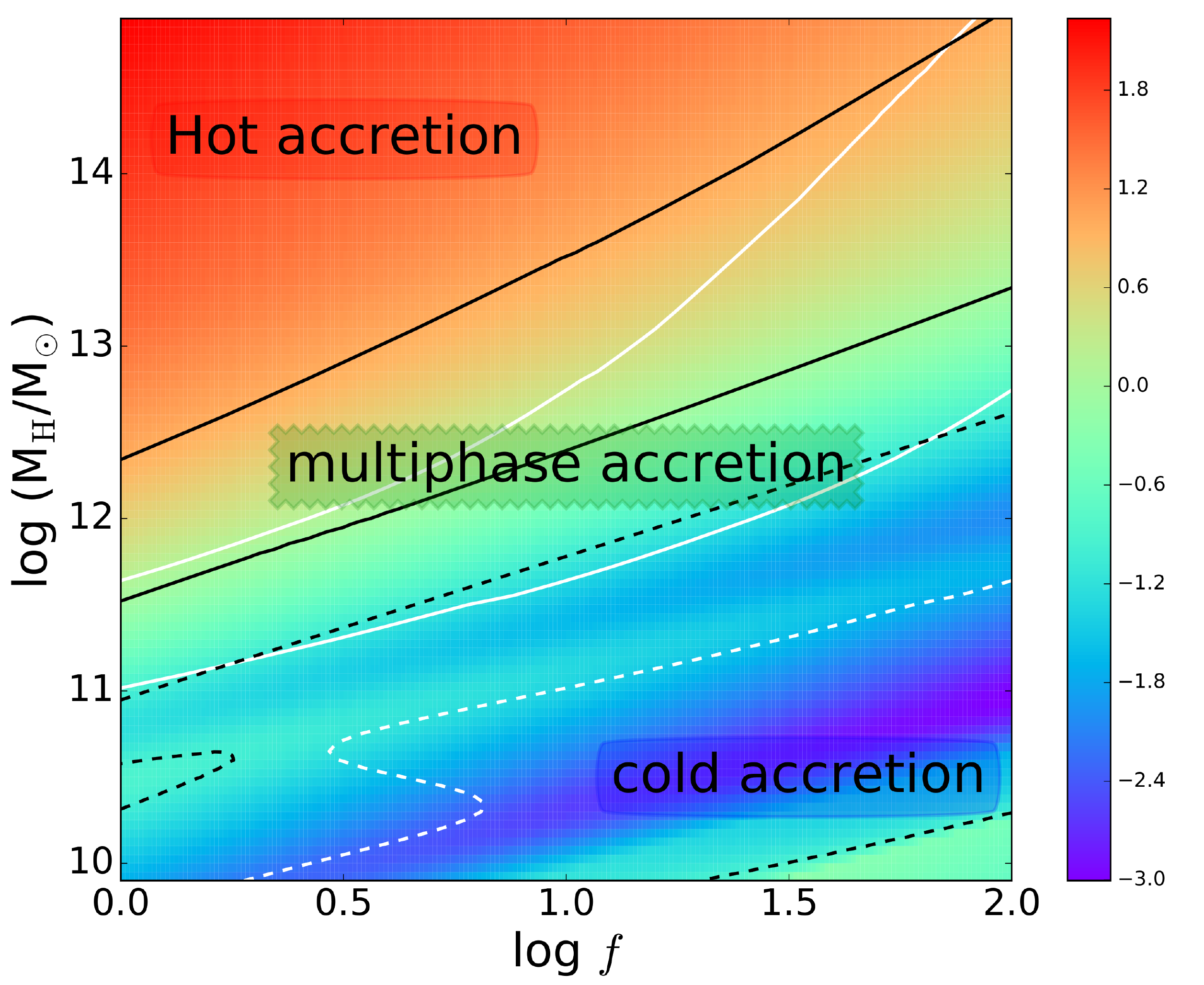}}
\caption{Ratio of the post-expansion cooling and halo dynamical timescales as a function of halo mass,
$M_\mathrm{H}$, and stream over-density, $f$, at redshift, $z$=2.  The colors represent the
log\,($t_\mathrm{cool,pe}/t_\mathrm{dyn, halo}$) (Eqs.~\ref{eqn:tcool} and \ref{eqn:tdynhalo})
whose corresponding values are indicated by the color bar. The solid black lines indicate 
log\,($t_\mathrm{cool,pe}/t_\mathrm{dyn, halo}$) of increasing values of 0 and 1 from
the middle to the top of the diagram.  Dashed black lines indicate
log\,($t_\mathrm{cool,pe}/t_\mathrm{dyn, halo}$)=$-$1. The white
contours indicate values of log\,($t_\mathrm{cool,ps}/t_\mathrm{expand}$) (Eqs.~\ref{eqn:texpand}
and \ref{eqn:tcool}) equal to 0 and 1 (solid), and $-$1 (dashed). For this analysis,
we have assumed log\,(r$_\mathrm{stream}$/r$_\mathrm{vir}$)=$-$1.5. We highlight regions
where the cooling times are significantly longer than the halo dynamical times
as ``hot accretion'', regions where the cooling time is significantly shorter than the
dynamical times as ``cold accretion'', and regions where the cooling times are of the same order-of-magnitude 
as the dynamical times as ``multiphase accretion'' (see text for details).  These
labels are illustrative and are not intended to precisely delineate a clean separation in the timescales.}
\label{fig:dyn_vs_cooling}
\end{figure}

To define the expansion time of the flow, we approximate this as the
inverse of the relative rate of change of pressure in the post-shock
gas as it expands.  Quantitatively, this is,

\begin{equation}
t_\mathrm{expand}=P_\mathrm{ps}/\dot{P}_\mathrm{ps}=-2\gamma r_\mathrm{stream}/\dot{r}_\mathrm{stream}\sim 2\gamma r_\mathrm{stream}/c_\mathrm{ps}
\label{eqn:texpand}
\end{equation}

\noindent
where $r_\mathrm{stream}$ is the initial radius of the stream before it
expands and $c_\mathrm{ps}$ is the sound speed of the gas immediately
after passing through the shock front.  As a first order approximation,
in the case of a homogeneous post-shock medium, one can easily compare
the cooling and expansion times.  Our definition of the expansion time
means it is a differential measure of the expansion and so the most
appropriate comparison is with the instantaneous cooling time of the
gas immediately after the shock.  The isobaric cooling time is defined as,

\begin{equation}
t_\mathrm{cool} (T) = T \left| \frac{dT}{dt} \right|^{-1}_P = \frac{5k_\mathrm{B} \rho  T}{2 \mu n_e n_{\rm H} \Lambda (T)} \approx 4.3 \times  \frac{5k_\mathrm{B}\mu m_\mathrm{p} T}{2\rho\Lambda (T)}
\label{eqn:tcool}
\end{equation}

\noindent
where $k_\mathrm{B}$ is the Boltzmann constant, $\Lambda\left(T\right)$
is the electronic cooling efficiency as a function of temperature,
$T$ \citep{sutherland93, gnat07}, $\mu$ is the mean molecular
weight\footnote{$\mu = 0.6 \, m_{\mathrm p}$ for a fully ionized gas.}, $m_\mathrm{p}$
is the mass of the proton, $\rho$ is the mass density of the cooling gas,
$n_e$ is the electronic density and $n_{\mathrm{H}}$ the Hydrogen particle
density\footnote{The factor 4.3 in Eq.~\ref{eqn:tcool} comes from
$n_e = 1.2 n_{\mathrm{H}}$ and $n_{\mathrm{H}} \approx \rho / (1.4 \, m_\mathrm{p})$.}.
Hereafter, the cooling time of the post-shock is denoted by, $t_\mathrm{cool, ps}$. We also refer to the cooling time after expansion, noted $t_{\rm cool, pe}$.

The other timescale with which to compare the cooling time is the dynamical
time of the halo. It is simple to estimate the dynamical time.
We estimate the dynamical time for matter falling from the virial radius
with a radial velocity of $v_{2}$ (the post-shock velocity) directed at
the center of the potential as,

\begin{equation}
t_\mathrm{ff}=\alpha \frac{r_\mathrm{vir}}{v_{2}}\approx t_\mathrm{dyn, halo}
\label{eqn:tdynhalo}
\end{equation}

\noindent
For this estimate, we assume a NFW dark matter potential \citep{navarro97}
and $\alpha$, is a factor of-order unity to account for the integrated
gravitational acceleration during the fall. Assuming that 
the characteristic velocity and radius are the virial values,
$v_\mathrm{vir}$ and $r_\mathrm{vir}$, the
dynamical time is independent of halo mass \citep{mo10}.

Fig.~\ref{fig:tscales_vs_MH} illustrates how these time scales
depend on the post-shock temperature, directly related to the halo mass (top x-axis) and for a range of
relative stream over-densities, $f$, and relative stream radii,
$r_\mathrm{stream}$/$r_\mathrm{vir}$ at z=2.
The initial stream penetrating the halo gas and the warm post-shock clouds are assumed to
have a temperature of 10$^4$ K, which is maintained through external
heating processes, such as the ionization by the meta-galactic flux or
ionizing photons from the galaxy in the halo. Since the post-shock
temperatures are proportional to velocity, the post-shock temperature
translates into dark matter halo mass. From this analysis, we see that the
expansion times are almost constant above $M_\mathrm{H}\!\ga\mathrm{few}
10^{10}\,\mathrm{M}_{\sun}$ since the virial shock is sufficiently
strong such that $t_\mathrm{expand}\propto t_\mathrm{dyn,halo}$.  As $f$
increases, the cooling times decrease systematically, which means for
low temperatures (low mass halos at z=2), the cooling time is always
less than the expansion time.  For very high mass halos, even for wide
streams with relatively high densities, the cooling time is much longer
than the expansion time and can be longer than the halo dynamical time
for low to moderately over-dense filaments. A wide range of values
in $f$ and $r_\mathrm{stream}$/$r_\mathrm{vir}$ lead to cooling times
that are less than the halo dynamical time and within an
order-of-magnitude of the expansion time.  Whenever the expansion time is
comparable to the cooling time, the gas flow will globally cool neither
purely adiabatically or isobarically. 

We can understand the competition between these time scales,
t$_\mathrm{cool, ps}$, t$_\mathrm{cool, pe}$, $t_\mathrm{dyn,halo}$,
and $t_\mathrm{expand}$, more clearly by
considering them as
a function of the stream over-density, $f$, and post-shock
temperature or halo mass (Fig.~\ref{fig:dyn_vs_cooling}).
For the case where $t_\mathrm{cool,ps}\!>>\!t_\mathrm{expand}$ and
$t_\mathrm{cool,pe}\!>\!t_\mathrm{dyn,halo}$, the post-shock gas will
expand rapidly remaining hot over at least a dynamical time of the
halo. In this regime the hot post-shock gas will likely mix with the
ambient halo gas preventing significant cooling. This is akin the
``hot mode accretion'' discussed by Dekel and collaborators (e.g.,
DB06).  In Fig.~\ref{fig:dyn_vs_cooling}, we have labeled this regime, "hot accretion". 
At the other extreme in Fig.~\ref{fig:dyn_vs_cooling},
when $t_\mathrm{cool,ps}\!<<\!t_\mathrm{expand}$ and
$t_\mathrm{cool,ps}\!<<\!t_\mathrm{dyn,halo}$, the gas cools before
significant expansion and before the stream has penetrated deeply
into the halo. Accretion in this regime corresponds to their ``cold
mode accretion'' and have labeled it as such in Fig.~\ref{fig:dyn_vs_cooling}.
In between these two regimes, $t_\mathrm{cool,pe}$,
$t_\mathrm{cool,ps}$, $t_\mathrm{dyn,halo}$, and $t_\mathrm{expand}$ are
all about the same order-of-magnitude. In this regime, there is not a
simple dichotomy between the types of accretion streams -- they can be
both hot and warm if the post-shock gas has a range of temperature and/or densities and
the thermodynamic evolution
of the post-shock gas will be more complicated. We labeled this regime
``multiphase accretion'' in Fig.~\ref{fig:dyn_vs_cooling}. 

\subsection{Inhomogeneous infalling streams and post-shock gas: differential
cooling}\label{subsec:diffcooling}

We now discuss
the physical mechanisms behind the development of multiphasic accretion streams.
Although our previous discussion of the relevant timescales considered 
homogeneous streams, density and velocity inhomogeneities may arise through 
the dynamics of the shock itself \citep{kornreich00, sutherland03}. 
Simulations show that stream are not accreted homogeneously but have substructure in both density
and velocity \citep{nelson16}. Inhomogeneities may arise due to a range of curvature in
accretion shock-fronts, translating into a range of Mach numbers and 
post-shock temperatures and densities. Density
fluctuations at constant shock velocity will also lead to inhomogeneities
in the post-shock gas in both temperature and density resulting in a range
of cooling times \citep{guillard09}. As we now discuss, once such ``differential cooling'' sets in, it acts like the thermal instability, leading to phase separation in the flow \citep{sharma12}, but over a finite range of time in the absence of a heating process balancing cooling of the hot phase.

To understand how the inhomogeneous post-shock gas evolves, we
investigate how small fluctuations in the density and/or temperature
are amplified. Neglecting the influence of any heating process, the
radiative cooling is not balanced and there
are no fixed equilibrium points. Following closely the development in
\citet{sharma12}, we begin with a parcel of gas that is over-dense relative
to an ambient medium. The magnitude of the over-density
is, $\delta\equiv |\delta \rho/\rho|\sim|\delta\,\mathrm{T}/\mathrm{T}|$
for isobaric conditions. The inverse of the effective cooling time of 
the over-dense gas parcel relative to the ambient medium is then
\begin{equation}
\frac{1}{t_\mathrm{cool,eff}} = -\frac{1}{t_\mathrm{cool,parcel}} + \frac{1}{t_\mathrm{cool,ambient}} \ .
\label{eqn:effcoolingdiff}
\end{equation}

\noindent
As already shown by \citet{sharma12}, for small over-densities, $\delta \leq$1, the inverse of the effective cooling time under isobaric conditions is
\begin{equation}
\frac{1}{t_\mathrm{cool,eff}} \sim -\delta \frac{\partial}{\partial\mathrm{\,\ln T}} \Bigg( \frac{1}{t_\mathrm{cool}} \Bigg)_\mathrm{P} = \frac{\delta}{t_\mathrm{cool}} \Bigg[ 2 - \frac{\diff\ln\Lambda(T)}{\diff\ln T} \Bigg] \ .
\label{eqn:effcoolingrate}
\end{equation}

Marginally denser gas cools faster
than the gas in which it is embedded and the cooling becomes a runaway
process for temperatures below a few 10$^5$\,K. We refer to this
 process as \textit{``differential cooling"}. 
As soon as heterogeneities form in approximate
pressure equilibrium with their surroundings, denser cooler regions
of the gas will cool rapidly, forming clouds, leaving a hotter, rarer,
high volume-filling inter-cloud medium. Thus, differential cooling leads
to multiphasic post-shock accretion flows. This
is akin to what occurs for thermally unstable gas \citep{field65} but the 
phase separation is transient in the absence of
a heating process of the hot gas balancing cooling.  

Of course, a natural consequence of differential cooling is
that even for halo masses where the post-shock gas has on average a long
cooling time, gas with fluctuations in temperature or
density may locally cool sufficiently rapidly to form a multiphased stream.

\begin{figure}[t]
\resizebox{\hsize}{!}{\includegraphics{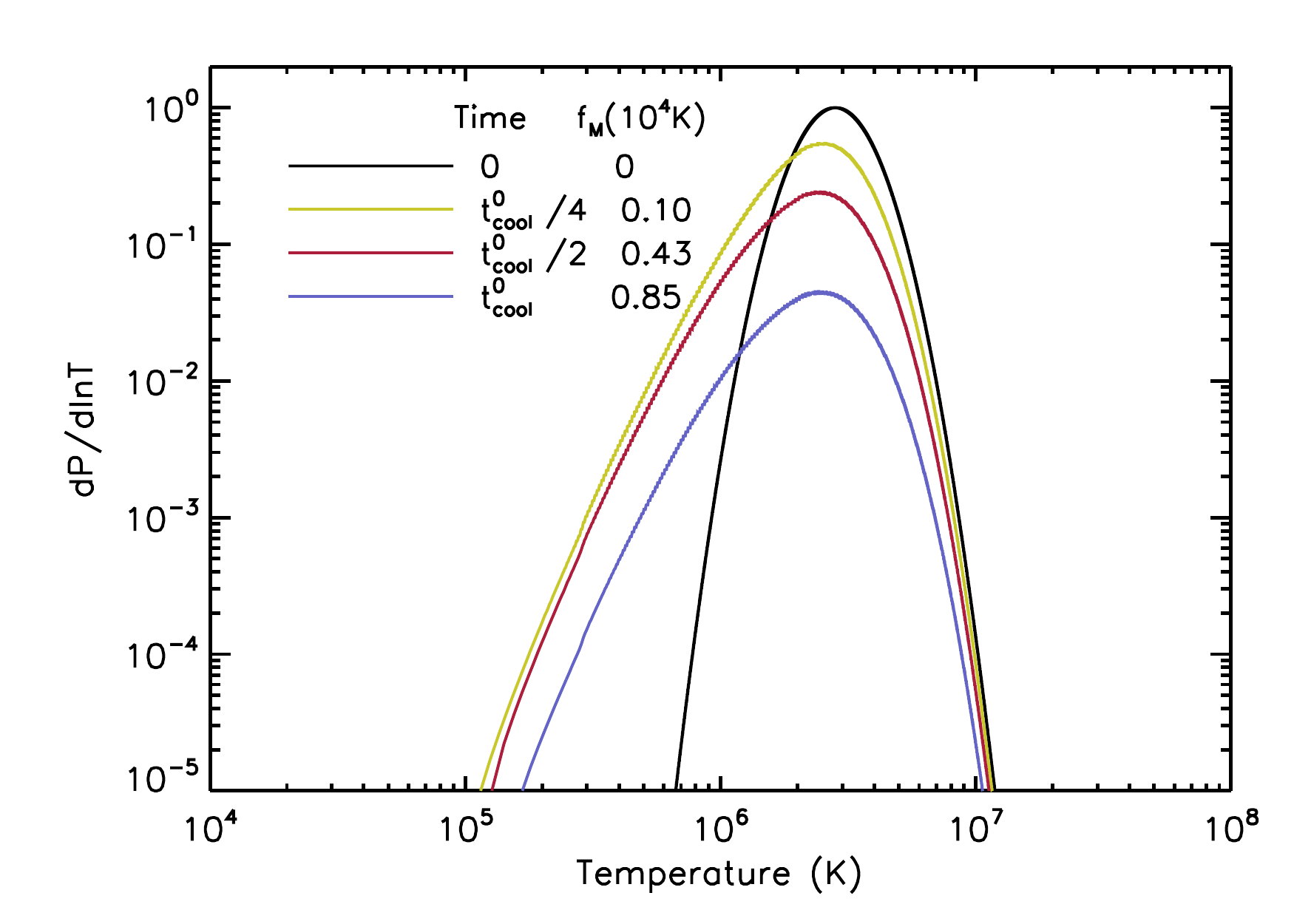}}
\caption{Probability distribution function of the gas temperature at different times
for isobaric cooling. The initial distribution (black line) is log-normal with a
dispersion, $\sigma_{\rm T}$ = 0.3, and an initial mean gas
temperature, $T_0$=2.7 $\times$ 10$^6$\,K. The distribution at a
time equal to the initial isobaric gas cooling time is plotted in blue,
that of half and a quarter of this time in
red and green, respectively. In the last column of the legend, we indicate the fraction of the gas that
has cooled to 10$^4$\,K within the indicated fraction of the cooling time.  For example,
in this illustration, after one cooling time, approximate 85\% of the gas has
cooled to 10$^4$\,K.}
\label{fig:cooling_pdf}
\end{figure}

Analytically, inhomogeneities can be accounted for as is done in phenomenological analyses of the
formation of stars through the use of probability distribution functions (PDFs) in gas density
and/or temperature \citep[e.g.,][]{hennebelle09}. 
We follow the thermodynamical evolution of the temperature and density PDFs using the energy equation, namely,
\begin{equation}
\dot{e} = -\rho \frac{\Lambda(T)}{4.3 \mu m_p} + P\,\frac{\dot{\rho}}{\rho^2} \ ,
\label{eqn:effcoolingrate}
\end{equation}
where $e$ is the specific internal energy.
We illustrate the isobaric evolution of inhomogeneities
in a flow using such an approach in Fig.~\ref{fig:cooling_pdf}. In this illustration,
we consider a log-normal PDF in temperature with a dispersion of $\sigma _ T = 0.3$ ($1\sigma$ dispersion of $\log \frac{\rho}{\rho _0} = \log \frac{T}{T_0}$, where $\rho _0$ and $T_0$ are the initial mean values of the density and temperature, with corresponding cooling time $t^0 _{\rm cool}$). We have chosen a gas pressure and mean
temperature appropriate for the post-shock gas of a flow into a
massive, 10$^{13}$ M$_{\sun}$, halo at z=2 (see \S\,\ref{sec:specificcase}). For this illustration, the gas
pressure is P/k = 5.8 $\times$ 10$^4$ K cm$^{-3}$, and the initial mean temperature $T_0$=2.7 $\times$ 10$^6$\,K. We find that a
substantial fraction of the accreting gas cools to 10$^4$\,K in a fraction of $t^0 _{\rm cool}$, while the peak of the hot gas temperature PDF does not shift with time (see Fig.~\ref{fig:cooling_pdf}). 

\begin{figure}[t]
\resizebox{\hsize}{!}{\includegraphics{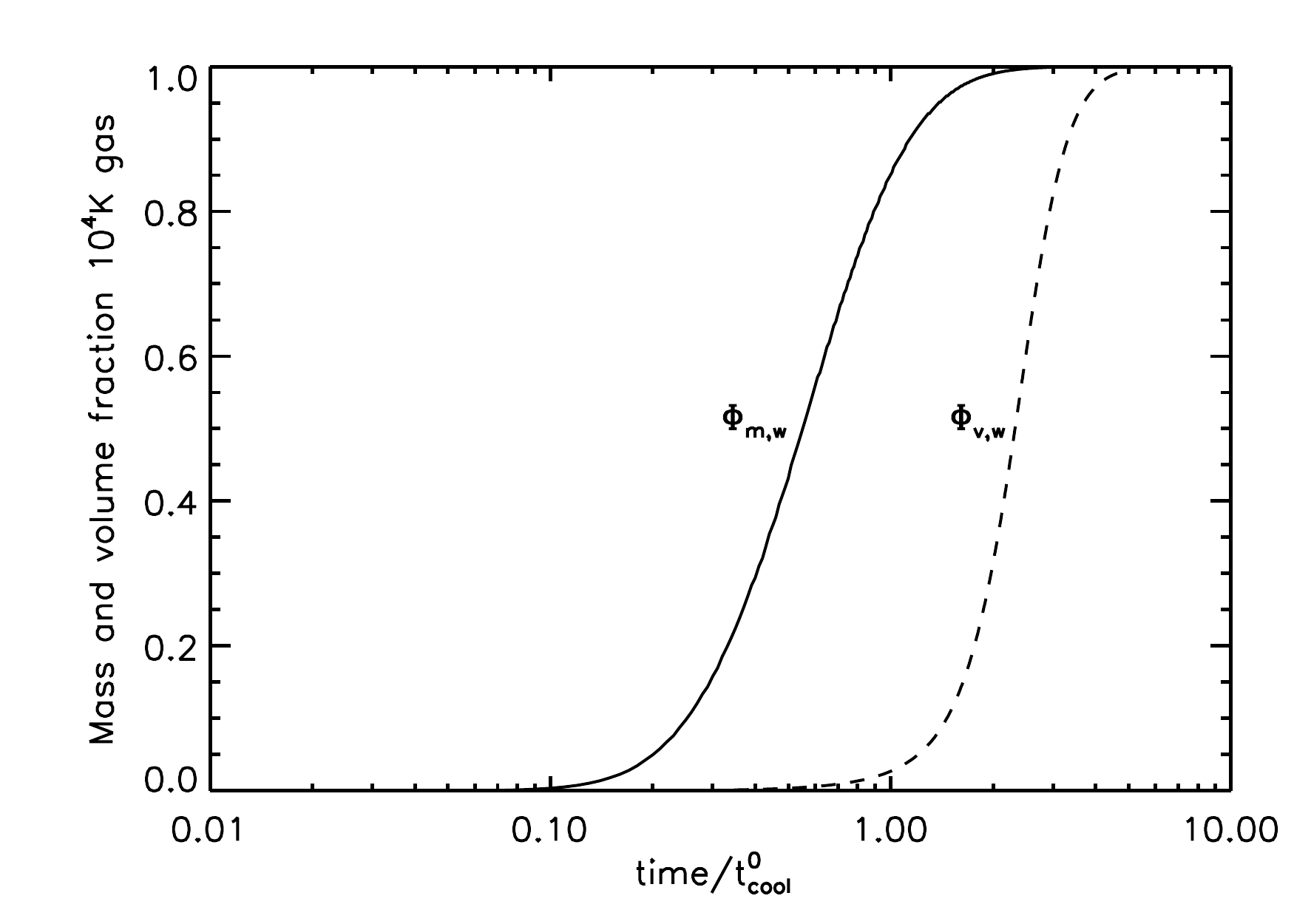}}
\caption{The change in the mass fraction, $\phi_\mathrm{m,w}$, and the volume
filling-factor, $\phi_\mathrm{v,w}$, of the warm, 10$^4$ K gas as a function of the
post-shock cooling time. The time scale is expressed in units of the isobaric
cooling time, $t^0 _{\mathrm cool}$ as defined in the text. The values of the pressure and
temperature are the same as those used in Fig.~\ref{fig:cooling_pdf}.}
\label{fig:cooling_mass_volume}
\end{figure}

We show in Fig.~\ref{fig:cooling_mass_volume} both the mass fraction and volume-filling factor
of the gas that cools to 10$^4$\,K as a function of time. Note that since the results shown on Figs.~\ref{fig:cooling_pdf} and \ref{fig:dyn_vs_cooling} are plotted as a function of the initial cooling time $t^0 _{\rm cool}$, to a first approximation they are not much dependent on the specific values of $\rho_0$ and $T_0$. Those results also apply to the gas cooling after expansion when $t_{\mathrm{expand}} << t^0 _{\rm cool}$. Figure~\ref{fig:dyn_vs_cooling} shows that the warm gas becomes the dominant mass phase, within about half of the post-shock cooling time, and that
the warm gas only fills a small fraction of the volume. For a larger value of the temperature dispersion $\sigma _{\rm T}$, $\Phi _{\rm m, w}$ increases over a broader range of fractional times distributed around the same mean value. 


In Fig.~\ref{fig:cooling_mass_volume}, the gas is multiphase for only a few $t^0 _{\rm cool}$.
The hot gas eventually cools because we have not considered any heating processes.
However, the process of fragmentation of the gas could be sustained if there is heating of the hot gas. Even in absence of heating,
after expansion the hot gas will eventually mix with the ambient halo gas thus sustaining the hot phase. Possible local sources of heating for the post-shock gas are the radiative
precursor of high velocity shocks, thermal conduction, radiation from the
surrounding hot medium, and dissipation of turbulence.
The mechanical energy input from active galactic nuclei (AGN), winds
generated by intense star formation, and other processes can plausibly
balance the cooling of the hot gas globally \citep[e.g.,][]{best07,
rafferty08}. Overall, there is more than enough energy, but what is
unknown is how and with what efficiency this energy is transferred to
the streaming gas. Even though simulations do not capture this process, it has
been suggested that the energy from the galaxy is transferred efficiently through
turbulent energy cascade and dissipation \citep{zhuravleva14, banerjee14}.

A poignant question to ask is can the flowing post-shock gas actually become
multiphase while cooling in a halo dynamical time?  It is a difficult
question to answer in the particular case of accretion streams because it depends on
how inhomogeneous the gas is. It will occur if the post-shock conditions are sufficiently inhomogeneous.
The gas will become multiphased around when $t_\mathrm{cool,pe}$ is the same order-of-magnitude as $t_\mathrm{dyn, halo}$. This justifies where
we placed the ``multiphase accretion'' label in Fig.~\ref{fig:dyn_vs_cooling}.
This figure shows that this occurs in the important mass range of M$_\mathrm{halo}$$\sim$10$^{11}$ to 10$^{13}$ M$_{\sun}$.  The values of course depend on the relative over-density of the streams ($f$) and through the characteristics of the halos, on redshift. Halos within this mass range is where the bulk of stars have formed in galaxies.
Moreover, because the volume filling factor of the warm gas is likely to be
always small, even in the cases where the mass fraction is large,
multiphase streams are likely to be difficult to identify observationally. We discuss
this further in \S\,\ref{sec:discussion}.

\section{Consequences of the formation of multiphased, cloudy accretion flows}
\label{sec:consequences}

\subsection{Turbulence in warm clouds}
\label{subsec:turbwarmclouds}

As the gas fragments, perhaps due to differential cooling and/or thermal
instabilities, we assume that some of the initial kinetic energy of
the stream is converted into turbulence within the warm component.
We parameterize the turbulence after phase separation as the ratio,
$\eta$, of the turbulent energy density of the clouds and the initial
bulk kinetic energy density of the stream.  We define this ratio as,

\begin{equation}
\label{eqn:eta}
\eta=\frac{\left<\rho_\mathrm{w}\right>_\mathrm{v} \sigma_\mathrm{turb}^2}{\rho_\mathrm{1} v_\mathrm{1}^2}
\end{equation}

\noindent where $\left<\rho_\mathrm{w}\right>_\mathrm{v}$
is the volume-averaged density of the warm clouds
($\left<\rho_\mathrm{w}\right>_\mathrm{v}$=$\phi_\mathrm{v,w}$
$\rho_\mathrm{w}$, where $\phi_\mathrm{v,w}$ is volume-filling factor
and $\rho_\mathrm{w}$ is the density of the warm clouds respectively),
$\sigma_\mathrm{turb}$ is the cloud-cloud velocity dispersion,
$\rho_\mathrm{1}$ and $v_\mathrm{1}$ are the pre-shock gas density and
velocity. We assume that $v_\mathrm{1}$ is equal to the virial velocity,
$v_\mathrm{vir}$. The initial density of the stream is related to the hot
halo density, $\rho_\mathrm{H}$, as $\rho_1\!=\!f\rho_\mathrm{H}$. The
amount of turbulence generated in the post-shock gas is likely determined by as yet poorly understood
and undoubtedly complex gas physics.
In our model we do not attempt to investigate the complexity involved in this
transformation of energy, we simply parameterize the amount of turbulence by
$\eta$ which we allow to be free (see \S\,\ref{subsec:cloudydisrupt}).

As we briefly alluded to in \S\,\ref{subsec:diffcooling}, as the turbulent energy dissipates,
it reheats the warm clouds,  moderating the cooling. The cooling of the clouds will also be moderated 
by mixing with the ambient halo gas as the stream penetrates deeper into
the halo potential \citep{fragile04}. 
Dissipation of turbulent kinetic energy in the hot phase may also contribute to
heating the post-shock gas as well as
the radiation from the high-speed accretion shock \citep{allen08}.
The parameter $\eta$ may be high enough to provide 
the required heating of the hot gas in the stream through turbulent dissipation and
mixing with the halo gas, perhaps instigating thermal instabilities. On the other hand, $\eta$ must
be low enough such that turbulent
dissipation and mixing does not prevent instabilities, both thermal and cooling, from growing
in the post-shock gas \citep{banerjee14, zhuravleva14}. These constraints and considerations
may ultimately provide limits on how much or how little turbulence is sustainable in the
post-shock gas, but determining these exact values requires further detailed study and will
not be considered here.

\subsection{Mass and momentum budget of the multiphase medium}
\label{massmomentumbudget}

We assume that as the stream flows into the halo, its mass flow rate is
conserved during the post-shock expansion and does not mix immediately with the
ambient halo gas.  This leads to the relation,

\begin{equation}
S\rho_2 v_2=\rho_\mathrm{1} v_\mathrm{1}
\label{eqn:Srho}
\end{equation}

\noindent
where $\rho_2$ is the density after the hot gas has expanded by a factor,
$S$, and similarly, $v_2$ is its velocity (Fig.~\ref{fig:sketch}).
The expansion factor, $S$, is defined as the ratio of the initial over
the final mass fluxes (per unit surface perpendicular to the flow). The
post-shock gas will ultimately reach pressure equilibrium with the
halo which implies, $\rho_\mathrm{w} T_\mathrm{w} = \rho_\mathrm{h}
T_\mathrm{h} = \rho_\mathrm{H} T_\mathrm{H}$, where $\rho_\mathrm{w}$
and $\rho_\mathrm{h}$ are the densities of the warm and hot components
and $T_\mathrm{H}$ is the temperature of the hot components after phase
separation in the post-shock gas.

\begin{figure*}[t]
\resizebox{\hsize}{!}{\includegraphics{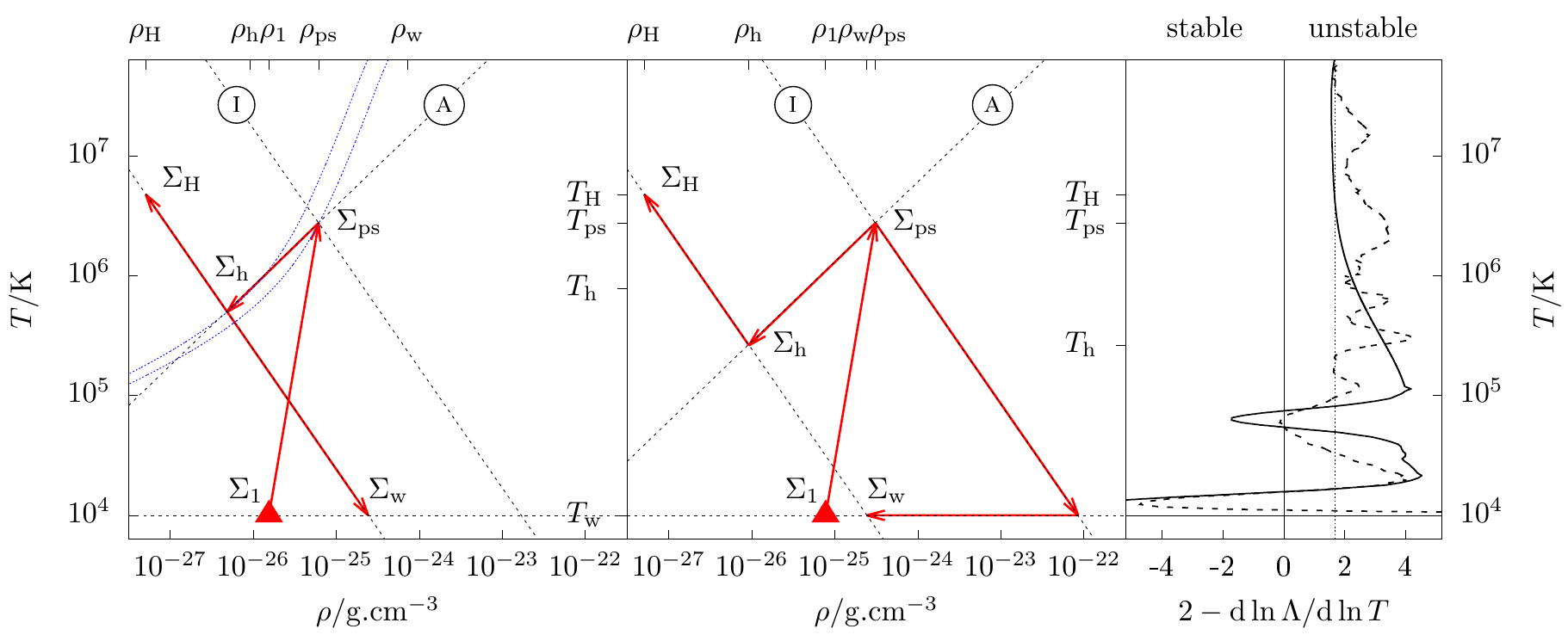}}
\caption{\textit{(left and middle)} Sketch of the thermodynamic path
for cases \textit{(1)} and \textit{(2)} (see text).  $\Sigma_1$ (red triangle)
indicates the density and temperature of the pre-shock gas. The gas is shocked reaching the point
$\Sigma_\mathrm{ps}$. Subsequently, the gas cools adiabatically due
to the expansion of the stream until the halo pressure is reached
at $\Sigma_\mathrm{h}$. Two phases separate. For case \textit{(1)},
phase separation occurs at $\Sigma_\mathrm{h}$, while for case \textit{(2)},
it occurs at $\Sigma_\mathrm{ps}$. In both cases, the hot component
mixes with the surrounding halo gas, reaching $\Sigma_\mathrm{H}$.
For case \textit{(1)}, the warm component cools radiatively and isobarically
to the point $\Sigma_\mathrm{w}$. For case \textit{(2)}, the same point is reached
but the gas takes a different thermodynamic path.  Dashed black lines
represent adiabats and isobars labeled A and I respectively. The two
blue-dashed curves in the left panel are contours of constant cooling
time, $t_\mathrm{cool}\sim0.2 t_\mathrm{dyn, halo}$ (upper curve) and
$t_\mathrm{cool}\sim0.12 t_\mathrm{dyn, halo}$ (lower curve). These curves
indicate that the cooling time during the expansion remains approximately constant.
\textit{(right)} Analysis of the isobaric differential cooling
``instability'' (Eq.~\ref{eqn:coolingcriteria}) of low-,
10$^{-3}$ solar (solid line) and solar-metallicity (dashed line) gas
as a function of temperature. When $2-\diff\ln\Lambda/\!\!\diff\ln T>0$
the gas can become heterogeneous through accelerated differential
cooling.}
\label{fig:T_rho_dLdT}
\end{figure*}

Before there is any momentum exchange with the halo gas, the momentum
of the streaming gas is conserved, implying,

\begin{equation}
S\left(\rho_2 v_2^2 + \eta\rho_1 v_1^2 + P_\mathrm{H}\right)=\rho_1
v_1^2 + P_1
\label{eqn:momconservation}
\end{equation}

\noindent where $P_1$ is the initial pressure of the stream.  We assume the
fragmenting gas radiates away its heat until reaching a floor temperature,
$T_\mathrm{w}$=10$^{4}$ K.  The hot component only cools adiabatically and
reversibly (no heat transfer and no entropy increase) expanding after it passed
through the shock front, namely, $\rho_\mathrm{h}^{\gamma}
P_\mathrm{ps}=\rho_\mathrm{ps}^{\gamma} P_\mathrm{H}$, where $\gamma$ is the
ratio of specific heats. This approximation holds until
the expanding gas reaches pressure equilibrium with the ambient halo gas.

The expansion factor, $S$, is derived from Eqs.~\ref{eqn:PH},
\ref{eqn:Srho}, and \ref{eqn:momconservation}, as,

\begin{equation}
	S=\frac{1}{2}\left(\eta+\frac{\gamma-1}{2f}\right)^{-1}
 \left(1-\sqrt{1-4\frac{\rho_\mathrm{H}}{\rho_{2}}
 \left(\eta f+\frac{\gamma-1}{2}\right)}\right)
\label{eqn:S}
\end{equation}

\noindent The expansion of the stream is important in our formulation.
It leads to the mixing of the expanding post-shock gas with the ambient
halo gas. This mixing couples the hot post-shock gas to the larger energy
reservoir of the ambient halo gas which acts as a thermostat preventing
the gas from cooling, thereby possibly maintaining the pressure necessary to
support a sustained shock.  As we discussed in \S\,\ref{sec:motivation},
it is the relative ratios of the thermal cooling time and the expansion
time that will influence how the stream evolves.

\subsection{Cloudy stream disruption}
\label{subsec:cloudydisrupt}

The relative cloud-cloud motions may lead to the warm clouds spreading beyond
the original radius of the stream.  So instead of the streams being highly
collimated as we assumed they are initially (before the virial shock),
the flows may de-collimate.  This can be thought of as disruption since
the warm clouds expand away from the original trajectory of the stream,
thus ``disrupting'' the flow. We define the timescale for disruption as
the cloud crossing time of the stream, namely,

\begin{equation}
t_\mathrm{disrupt}=\frac{r_\mathrm{stream}}{\sigma_\mathrm{turb}}
\label{eqn:tdisrupt}
\end{equation}

\noindent
$\sigma_\mathrm{turb}$ may be computed from $\eta$, $f$, and the
volume-filling factor of the warm clouds, $\phi_\mathrm{v,w}$.
From Eqs.~\ref{eqn:PH} and \ref{eqn:eta}, assuming pressure equilibrium
between gas phases, yields,

\begin{equation}
\begin{aligned}
\sigma_\mathrm{turb}^2&= v_\mathrm{1}^2 (\eta f/\phi_\mathrm{v,w})
	(T_\mathrm{w}/T_\mathrm{H}) \\
	&= \frac{2}{(\gamma-1)}\left(\frac{k_\mathrm{B}}{\mu m_p}\right)(\eta f/\phi_\mathrm{v,w}) T_\mathrm{w}
\end{aligned}
\label{eqn:sigma_turb}
\end{equation}

\noindent
Characterized this way, $\sigma_\mathrm{turb}^2/v_\mathrm{1}^2>\eta$. For
a wide range of relative amounts of turbulent energy, stream over-density,
and volume-filling factor of the warm gas, the cloud-cloud dispersion
be up to $v_\mathrm{vir}$/2. The dynamical evolution of the stream is determined
by the ratio of the disruption
and halo dynamical time.  If $t_\mathrm{disrupt} \gg t_\mathrm{dyn,
halo}$, then the warm clouds within the stream will remain collimated
as they flow, otherwise, the streams will de-collimate.

\section{A specific case: $10^{13}\,\mathrm{M}_{\sun}$ halo at $z$=2}
\label{sec:specificcase}

To gauge whether any of the phenomenology we have discussed is pertinent
astrophysically, we calculate the stream characteristics for a single
dark matter halo of mass, 10$^{13}$\,M$_{\sun}$, at redshift 2.
We chose this halo mass and redshift because in DB06, halos at this mass and redshift
were determined to have substantial accretion rates in the ``hot mode''. Our analysis in
\S\,\ref{sec:motivation} indicated that this halo mass and redshift would be a revealing
illustration of how the impact of expansion and accelerated differential
cooling might change the physical characteristics of gas accretion flows (i.e., it would no 
longer simply be ``hot mode accretion''). 
Both the halo mass and the redshift set the initial stream density and
the characteristics of the halo gas.  We provide all the characteristics
of the halos and initial gas conditions in the Appendix.

\subsection{Why are streams cloudy?}\label{subsec:biphasic}
\label{whycloudy}

There are two limiting cases to specifically consider when attempting to
understand the development of a biphasic stream.  The two cases are:
\textit{(1)} $t_\mathrm{expand} \ll t_\mathrm{cool, ps}$ and \textit{(2)}
$t_\mathrm{expand} \gg t_\mathrm{cool, ps}$.  In case \textit{(1)}, the warm
phase develops only after the expansion has occurred, while in case
\textit{(2)} the clouds form before the expansion.  We sketch the thermodynamic
evolution (``path'') of a stream in Fig.~\ref{fig:T_rho_dLdT}.  To make these
illustrations, we adopted $f=30$ and $r_\mathrm{stream}$/$r_\mathrm{vir}$=0.01
for case \textit{(1)}, and $f=150$ and $r_\mathrm{stream}$/$r_\mathrm{vir}$=0.1
for case \textit{(2)}.

In case \textit{(1)}, clouds fill the entire expanse of the expanded flow.
Adiabatic expansion occurs before radiative cooling becomes important,
the stream cools and the density of the hot phase declines without a
change in entropy.  In case \textit{(2)}, the two phase separate before
expansion. In both cases, the hot post-shock gas reaches pressure
equilibrium and mixes with the ambient halo gas.

\begin{figure}[t]
\resizebox{\hsize}{!}{\includegraphics{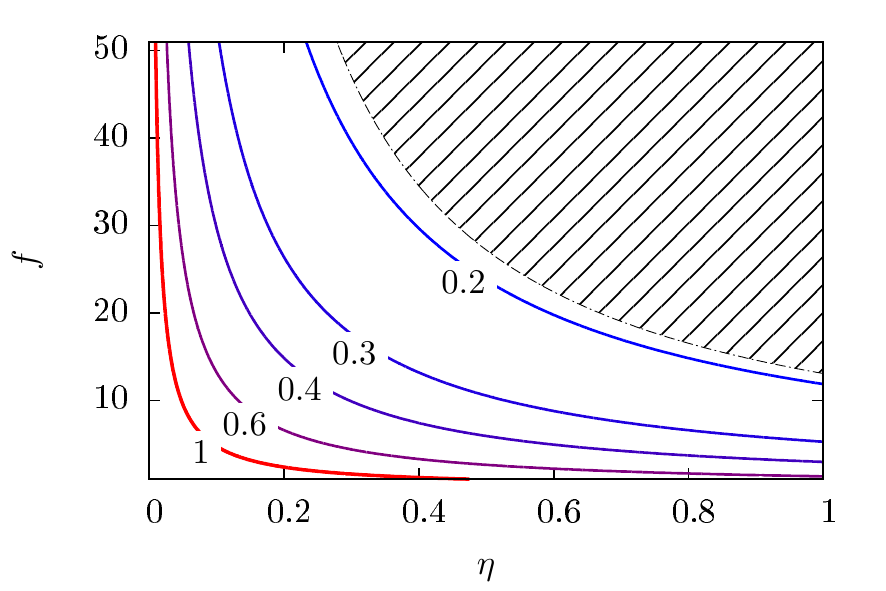}}
\caption{Disruption of the flow as a function of the filament
over-density, $f$, and level of turbulence, $\eta$. The contours represent
constant ratios of $t_\mathrm{disrupt}/t_\mathrm{dyn, halo}$ as
labeled (cf. Eqs.~\ref{eqn:tdisrupt} and \ref{eqn:tdynhalo}).
We assume a volume-filling factor of 0.1 (see Fig.~\ref{fig:cooling_mass_volume})
for the warm clouds and
$r_\mathrm{stream}=r_\mathrm{vir}$/10.  In regions with values less
than 1, the streams are ``disrupted'' (\S\,\ref{subsec:disrupt}). The
shaded region indicates regions that are forbidden because for these
values of the parameters, the post-shock pressure is less than the
halo pressure.  We note that because $t_\mathrm{disrupt}\propto
\sigma_\mathrm{turb}^{-1}$, the contours of constant
$t_\mathrm{disrupt}/t_\mathrm{dyn, halo}$ are shaped like contours of
constant $\sigma_\mathrm{turb}$ in the same plane. For example, the
contour, $t_\mathrm{disrupt}/t_\mathrm{dyn, halo}$=0.2, is close to the
contour for $\sigma_\mathrm{turb}$=200~km~s$^{-1}$, which is almost half
the initial velocity of the flow. }
\label{fig:disruption1} 
\end{figure}

These two conditions are of course for the extreme cases, in reality,
the gas will have $t_\mathrm{expand} \sim t_\mathrm{cool, ps}$ (Fig.~\ref{fig:dyn_vs_cooling}). For
these intermediate cases, the thermodynamic evolution is more complex,
but the clouds reach the same final thermodynamical state.  The cooling
length at the post-shock temperature is the size over which structures
can cool isobarically (i.e., cooling length,
$\lambda_\mathrm{cooling, ps}\!=\!c_\mathrm{ps}\,t_\mathrm{cool, ps}$).
In case \textit{(1)}, the cooling length is much larger than the stream radius,
$\lambda_\mathrm{cooling, ps}\!>>\!r_\mathrm{stream}$, clouds may form over all scales
within the stream.  In case \textit{(2)}, it is much smaller,
$\lambda_\mathrm{cooling, ps}\!<<\!r_\mathrm{stream}$, and
clouds may form over a range of sizes smaller than the stream radius.
The expansion of the hot gas does not inhibit the growth of thermal and differential cooling 
instabilities because the decrease in the pressure is roughly compensated
by the decrease of temperature along an adiabat in the expression of
the cooling time. In other words, for the post-shock temperature, i.e.,
for the halo mass we have adopted, the gas cooling time remains roughly
constant as the gas expands (Fig.~\ref{fig:T_rho_dLdT};
see also Fig.~\ref{fig:tscales_vs_MH}). Eventually,
the warm phase equilibrates at approximately the halo pressure and
the hot phase mixes with the halo gas.  We are obviously considering
fragmentation on scales much smaller than the scale height of the
gravitational potential well and thus we can safely ignore thermal
stabilization by convection \citep{balbus89, sharma10}.

\subsection{Does the accretion flow disrupt?}
\label{subsec:disrupt}

\begin{figure}[t]
\resizebox{\hsize}{!}{\includegraphics{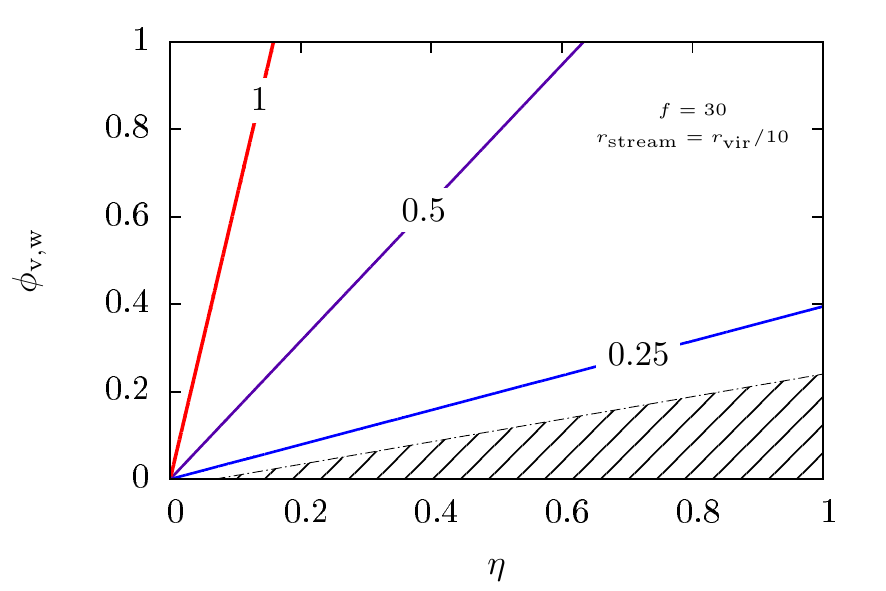}}
\caption{Disruption of the flow as a function of the volume-filling
factor of the warm gas and level of turbulence. The contours represent
constant ratios of t$_\mathrm{disrupt}$/t$_\mathrm{dyn, halo}$ as labelled
(cf. Eqs.~\ref{eqn:tdisrupt} and \ref{eqn:tdynhalo}). We assume $f$=30
and r$_\mathrm{stream}=r_\mathrm{vir}$/10.  In regions with values less
than 1, the streams are disrupted. The shaded region has the same meaning
as in Fig.~\ref{fig:disruption1}.}
\label{fig:disruption2}
\end{figure}

The only process we consider in determining whether or not the
warm clouds will travel coherently towards the galaxy proper
as observed in numerical simulations \citep[e.g.][]{brooks09,
danovich15} is the cloud-cloud velocity dispersion.  The cloud-cloud
dispersion will broaden the stream as it penetrates into the halo.
Fig.~\ref{fig:disruption1} and Fig.~\ref{fig:disruption2} show
contours of $t_\mathrm{disrupt}/t_\mathrm{dyn, halo}$ for a constant
$r_\mathrm{stream}$ (Eqs.~\ref{eqn:tdisrupt} and \ref{eqn:tdynhalo}).
We find that the disruption time is shorter than or approximately equal
to the halo dynamical time.  Thus it appears that for a wide range
of relative turbulent energy densities, stream over-densities, and
volume-filling factors of the warm gas, the flows will not simply fall
directly into the potential as a highly collimated, coherent streams.
In reality, the clouds are dynamical entities, we expect clouds to keep
forming through cooling as other clouds are destroyed
by hydrodynamic instabilities and heated by dissipation \citep{cooper09}.

\section{Discussion}
\label{sec:discussion}

We now discuss broadly how our findings relate to aspects of galaxy
formation and evolution.

\subsection{Are virial shocks persistent?}
\label{subsec:shockstability}

In our scenario, the existence of a hot phase in hydrostatic
equilibrium supports a persistent shock \citep{binney77, maller04}.
Cosmological simulations appear to show a similar phenomenology as that
described in DB06.  Depending on the mass of the halo and redshift,
streams penetrate at about one to many 100s of $\mathrm{km\,s^{-1}}$
\citep{vandeVoort12, goerdt15a} or much greater than the sound
speed of the stream, c$_\mathrm{s}$$\sim$10-20 $\mathrm{km\,s^{-1}}$.
Simulated accretion shocks are ``isothermal'' at high Mach numbers and
not stable \citep[e.g.,][]{nelson16, nelson15}. Perhaps this uniform
isothermality is due to the spatial and temporal resolutions adopted
in the cosmological simulations. The scales that simulations must
probe are roughly delineated in one-dimensional shock calculations.
\citet{raymond79} show that atomic shocks with velocities of $\sim$100
$\mathrm{km\,s^{-1}}$ reach their post-shock temperatures within a
distance of $\approx$1-10 $\times$10$^{15}$ cm in less than $\sim$30 yrs.
For higher shock velocities, the spatial and temporal scales will be even
shorter \citep{allen08}.  The gas cools after the shock on timescales that
are at most only a couple of orders-of-magnitude longer. In addition,
in order to capture the differential cooling and thermal instabilities
in the post-shock gas, resolutions much finer than the Field length are
required \citep{koyama04, gressel09}. Simulations should be specifically
designed to capture the multiphase nature of streams penetrating halos to
test our scenario by resolving the Field length \citep{koyama04}. Since
the Field length decreases strongly with decreasing temperature, this
is most easily done with \textit{ad hoc} floor temperature higher than 10$^4$
K but lower than the virial temperature as done for simulating thermal
instabilities in cool core clusters \citep{mcCourt12}.

The resolution and temporal scales necessary to resolve high Mach
number shocks are not achievable in galaxy- or cosmological-scale
simulations. To overcome this limitation, numericists use artificial
viscosity in the form of a dissipative term either in dynamical equations
or dispersion relations, depending on the properties of the gas or
the flow \citep[e.g.,][]{kritsuk11,price12, hu14, beck16}. Artificial
viscosity spreads the shock over several resolution elements enabling
simulations to resolve heating and cooling across the shock front.
The Reynolds number is inversely proportional to the kinetic viscosity
of the fluid. If the flow properties are unchanged but the viscosity
increased, the Reynolds number of the flow will be artificially
low. Simulations with artificial viscosity have flows with low Reynolds
numbers.  Simulated low $\mathrm{Re}$ flows, $\mathrm{Re}\la1000$, tend
to be laminar. Those with low spatial and temporal resolutions, due to not
resolving the Field length and having unrealistically low Reynolds numbers
likely fail to produce biphasic turbulent flows \citep[see, e.g.,][
for discussion]{kritsuk02, sutherland03, koyama04, kritsuk11, nelson16}.

\subsection{Nature of Flows into Galaxies: Observational tests}
\label{subsec:natureofflows}

As a consequence of our assumption that all energetic quantities scale
as $v_\mathrm{vir}$, the cloud-cloud dispersion is a simple linear
function of $\eta$, $f$, and $\phi_\mathrm{v,w}^{-1}$.  This relation
implies that turbulent velocities of the warm clouds in the post-shock
gas are independent of both halo mass and redshift. In principle
this means that post-shock streams may be turbulent in any halo at any
redshift. The reality is probably much more complex, through both macro- and microscopic
gas physics which is not yet well-understood,
the 2 parameters, $\eta$ and $\phi_\mathrm{v,w}$, likely depend on
the accretion velocity and the physical state of the ambient halo gas --
both of which undoubtedly depend on redshift and halo mass.

Our scenario has observationally identifiable consequences. If our
scenario is realistic, then observations should reveal:  \textit{(1)}
clumpy, turbulent streams; \textit{(2)} strong signs of the dissipation of
turbulent mechanical energy in the warm medium \citep[e.g.,][]{guillard09,
ogle10, tumlinson11}. The situation described in our model where a large fraction of the bulk 
kinetic energy of the accretion flow is transfered to turbulent motions 
amongst cold clouds is observed in large-scale galaxy colliding flows, like the
situation in the Taffy Galaxies or the Stephan's Quintet, where
we see evidence for this energy cascade \citep{peterson12, cluver10}. 
In Stephan's Quintet, two atomic gas filaments 
are colliding at $\sim$1000 $\mathrm{km\,s^{-1}}$ and yet instead of
finding intense X-ray emission from the post-shock gas, most of the
bulk kinetic energy is contained in the turbulent energy of the warm
molecular gas \citep{guillard09, guillard10}. Remarkably, roughly 90\%
of the bulk kinetic energy has not been dissipated in the large-scale
shock and is available to drive turbulence. If gas in halos is multiphase 
and turbulent, then it may be the case as well 
for accretion streams. 

Obviously, a clumpy stream is difficult to identify
as such through absorption line spectroscopy and this may explain why
streams have not been conspicuously identified so far.  This is most obviously
seen in Fig.~\ref{fig:cooling_mass_volume} where despite a large fraction of
the gas is warm, its volume-filling factor is minuscule. Along most
lines-of-sight, absorption spectroscopy is expected to sample only the
hot, high volume-filling factor halo gas or probe a population of warm
ambient halo clouds \citep{maller04, tumlinson11, werk14}. The clouds
should be looked for in emission. Their emission can be powered by the
UV radiation from the galaxy but also through the localized dissipation
of turbulent energy and through losses of their gravitational potential
energy as they fall into the halo.  This may have already been observed
in Ly$\alpha$ \citep[e.g.,][]{cantalupo14, martin15}.  In particular,
it would be promising to interpret spectral-imaging observations such as
those provided by MUSE on the ESO/VLT within the context of our scenario
\citep[see][]{borisova16, fumagalli16, vernet17}.

\subsection{Moderating the accretion rate: Biphasic streams and increased
coupling between ``feedback'' and accretion}
\label{subsec:feedbackandaccretion}

In our phenomenological model, two mechanisms moderate the accretion
efficiency on to galaxies: \textit{(1)} disruption and fragmentation of
the flow; \textit{(2)} interaction between streams and outflows of mass,
energy, and momentum due to processes occurring within galaxies (e.g.,
AGN, intense star-formation).

First, streams potentially become multiphase and turbulent leading to short disruption
times resulting in de-collimation. Any de-collimation undoubtedly leads
to longer accretion times and thus lower overall accretion efficiencies
compared to smooth isothermal streams \citep{danovich15, nelson16}.
The post shock gas becomes multiphase over a wide range of halo masses at z=2.
A fraction of the initial stream
mass flow becomes hot gas and ultimately mixes with the surrounding
ambient hot halo gas.  Thus even accretion streams potentially feed gas into the hot halo
which may have long cooling times compared to the halo free fall time
\citep{white78, maller04}.  

Second, simulations indicate that the mass and energy outflows from
galaxies can interact with streams, regulating or even stopping the
flow of gas \citep[e.g.,][]{ferrara05, dubois13, nelson15, lu15}.
Simulated streams are relatively narrow \citep[e.g.,][]{ocvirk08,
nelson13, nelson16} and generally penetrate the halo perpendicular
to the spin axes of disk galaxies and their directions are relatively
stable for long periods \citep[e.g.,][]{pichon11, dubois14, welker14,
laigle15, codis15, tillson15}.  Feedback due to mechanical and radiative
output of intense galactic star formation and active galactic nuclei
is observed to be highly collimated in inner regions of disk galaxies
\citep[opening angle, $\Omega\!\!\sim\!\!\!\pi$ sr, e.g.,][]{heckman90,
lehnert95, lehnert96, beirao15}. In the case of dwarf galaxies, their
outflows are generally more weakly collimated \citep[e.g.,][]{marlowe95,
marlowe97, martin98, martin05}. The geometry of the accretion flows and
the significant collimation of outflows from galaxies in simulations,
result in only weak direct stream-outflow interactions.  However,
accretion flows in simulations can be moderated or stopped when the
halo gas is pumped with mass and energy via feedback to sufficiently
high thermal pressures and low halo-stream density contrasts to
induce instabilities in the stream and disrupt it; or when the halo gas
develops a sufficiently high ram pressure in the inner halo due to angular
momentum exchange between the gas, dark matter, and galaxy disk to disrupt
accretion flows \citep{vandeVoort11, dubois13, nelson16}.

The processes we have described are generic to flows, whether they are
inflows or outflows.  It is only the context and timescales that change
\citep{thompson16}.  Just as with the accretion flows modeled here, we
also expect the galaxy winds to be highly uncollimated as they flow from
the galaxy due to the formation (and destruction) of turbulent clouds.
Thus, the generation of turbulent cloudy media in both accretion flows and
starburst-driven outflows will allow for efficient interaction between
these two types of flows. This dynamical interaction likely sustains
turbulence in the halo compensating for the dissipation. It is perhaps
through this interaction that galaxies become ``self-regulating'' on a
halo scale \citep[e.g.,][]{fraternali13}, and not only on a galaxy scale
\citep[e.g.,][]{L13, lehnert15}.

While the fate of the turbulent clouds is beyond the scope this paper,
the qualitative implication is that the gas accretion efficiency may
be moderated through the generation of turbulence in biphasic flows.
The fragmented, turbulent nature of the gas in streams and outflows likely
makes their dynamical and thermal interaction and coupling efficient. 
Note that other mechanisms, like the growth of Rayleigh–Taylor and Kelvin-Helmholtz instabilities, associated with gas cooling, can also trigger the formation of cold clouds in the surrounding halo \citep[e.g.][]{keres09c}.
Moderating the overall gas accretion efficiency onto galaxies may help
to alleviate two significant challenges in contemporary astrophysics:
the distribution of the ratio of the baryonic to total halo mass as
a function of halo mass \citep[e.g.][]{behroozi13}, where low mass
galaxies have especially low baryon fractions; and the requirement for
models to drive extremely massive and efficient outflows to reduce the
baryon content of galaxies \citep[e.g.,][]{hopkins12, hopkins16}.

\section{Conclusions}
\label{sec:conclusions}

We developed a phenomenological model of filamentary gas accretion,
``streams'', into dark matter halos. We assume both that streams penetrate ambient
hot halo gas as homogeneous flows of 10$^4$ K gas and that they undergo a shock
at the virial radius of the halo. The
ingredients of the model, those which sets it apart from other phenomenological
models of gas accretion, are that we assume the ``virial shock''
is sustained, the post-shock gas expands into a ambient hot halo gas, and through
several mechanisms or characteristics of the shock front, the post-shock gas is inhomogeneous.
To gauge whether this model is astrophysically pertinent, we discuss the
thermodynamic evolution of a single stream penetrating a dark matter halo
of mass 10$^{13}$ M$_{\sun}$ at z=2. From this analysis, we find that:

\begin{itemize}

\item The post-shock gas expands into the halo gas and it can fragment due
to differential cooling and hydrodynamic instabilities.  
Instabilities lead to the formation of a
biphasic flow.   It is the formation of a hot post-shock phase
which mixes with the ambient hot halo gas, ultimately limiting how much of the gas can
cool. As a result of the phase separation and the pressure provided by the
hot post-shock gas, we argue that the virial shock may be sustained. However, we have not
analyzed the sustainability of the shock in detail in this paper.

\item The development of a biphasic medium
converts some of the bulk kinetic energy into random
turbulent motions in the gas \citep[e.g.,][]{hennebelle99, kritsuk02}.
The turbulent energy cascades from large to small scales and across
gas phases.  The flows, while retaining significant bulk momentum as
they penetrate into the halo, are turbulent with cloud-cloud dispersion
velocities that can be up to 1/2 of the initial velocity of the stream.

\item For a wide range of turbulent energy densities, our model shows
that the stream will lose coherence in less than a halo dynamical time.
We emphasize that the turbulent energy density is not in reality a free
parameter but is determined by macro- and microscopic multiphase gas physics about which
we have only a rudimentary understanding. To understand
what processes regulate the amount of turbulence in streams, high
resolution simulations of accreting gas need to be made and additional
multi-wavelength observations useful for constraining the properties of
turbulent astrophysical flows are necessary.

\end{itemize}

The post-virial shock gas is not isothermal, accretion streams are
both hot and cold. The ``hot-cold dichotomy'' (see DB06)
is no longer a simple function of whether or not the shock is stable,
but now relies both on the shock occurring and under what circumstances
the post-shock gas becomes multiphase and turbulent.
However, we have discussed may apply if there is no virial shock provided that inflowing
gas is hot and already inhomogeneous \citep{kang05,cen06}.
Thus, even in absence of a virial shock, the gas
may become multiphased by compression as it falls deeper into the halo
potential.

Moderating the gas accretion efficiencies on to galaxies through this
and other mechanisms may help to alleviate some significant challenges
in theoretical astrophysics.  If gas accretion is actually not highly
efficient, then perhaps models will no longer have to rely on highly
mass-loaded outflows to regulate the gas content of galaxies.  It is
likely that the underlying physical mechanisms for regulating the
mass flow rates and evolution of outflows are very similar to those
that regulate gas accretion \citep{thompson16}.  If so, then observing
outflows in detail can provide additional constraints on the physics of
astrophysical flows generally. We do not only have to rely on apparently
challenging detections of direct accretion onto galaxies.

\begin{acknowledgements}
This work is supported by a grant from the R\'egion Ile-de France.
NC wishes to thank the DIM ACAV for its generous support of his thesis
work. We thank Yohan Dubois for insightful discussions and an anonymous referee for
their skepticism, poignant questions, and significant challenges which helped to
improve our paper significantly.
\end{acknowledgements}

\bibliographystyle{aa}
\bibliography{accretionrefs.bib}

\begin{thebibliography}{111}
\expandafter\ifx\csname natexlab\endcsname\relax\def\natexlab#1{#1}\fi

\bibitem[{{Alatalo} {et~al.}(2015){Alatalo}, {Appleton}, {Lisenfeld},
  {Bitsakis}, {Lanz}, {Lacy}, {Charmandaris}, {Cluver}, {Dopita}, {Guillard},
  {Jarrett}, {Kewley}, {Nyland}, {Ogle}, {Rasmussen}, {Rich},
  {Verdes-Montenegro}, {Xu}, \& {Yun}}]{alatalo15}
{Alatalo}, K., {Appleton}, P.~N., {Lisenfeld}, U., {et~al.} 2015, \apj, 812,
  117

\bibitem[{{Allen} {et~al.}(2008){Allen}, {Groves}, {Dopita}, {Sutherland}, \&
  {Kewley}}]{allen08}
{Allen}, M.~G., {Groves}, B.~A., {Dopita}, M.~A., {Sutherland}, R.~S., \&
  {Kewley}, L.~J. 2008, \apjs, 178, 20

\bibitem[{{Anderson} \& {Bregman}(2010)}]{anderson10}
{Anderson}, M.~E. \& {Bregman}, J.~N. 2010, \apj, 714, 320

\bibitem[{{Appleton} {et~al.}(2013){Appleton}, {Guillard}, {Boulanger},
  {Cluver}, {Ogle}, {Falgarone}, {Pineau des For{\^e}ts}, {O'Sullivan}, {Duc},
  {Gallagher}, {Gao}, {Jarrett}, {Konstantopoulos}, {Lisenfeld}, {Lord}, {Lu},
  {Peterson}, {Struck}, {Sturm}, {Tuffs}, {Valchanov}, {van der Werf}, \&
  {Xu}}]{appleton13}
{Appleton}, P.~N., {Guillard}, P., {Boulanger}, F., {et~al.} 2013, \apj, 777,
  66

\bibitem[{{Balbus} \& {Soker}(1989)}]{balbus89}
{Balbus}, S.~A. \& {Soker}, N. 1989, \apj, 341, 611

\bibitem[{{Banerjee} \& {Sharma}(2014)}]{banerjee14}
{Banerjee}, N. \& {Sharma}, P. 2014, \mnras, 443, 687

\bibitem[{{Beck} {et~al.}(2016){Beck}, {Murante}, {Arth}, {Remus}, {Teklu},
  {Donnert}, {Planelles}, {Beck}, {F{\"o}rster}, {Imgrund}, {Dolag}, \&
  {Borgani}}]{beck16}
{Beck}, A.~M., {Murante}, G., {Arth}, A., {et~al.} 2016, \mnras, 455, 2110

\bibitem[{{Behroozi} {et~al.}(2013){Behroozi}, {Wechsler}, \&
  {Conroy}}]{behroozi13}
{Behroozi}, P.~S., {Wechsler}, R.~H., \& {Conroy}, C. 2013, \apj, 770, 57

\bibitem[{{Beir{\~a}o} {et~al.}(2015){Beir{\~a}o}, {Armus}, {Lehnert},
  {Guillard}, {Heckman}, {Draine}, {Hollenbach}, {Walter}, {Sheth}, {Smith},
  {Shopbell}, {Boulanger}, {Surace}, {Hoopes}, \& {Engelbracht}}]{beirao15}
{Beir{\~a}o}, P., {Armus}, L., {Lehnert}, M.~D., {et~al.} 2015, \mnras, 451,
  2640

\bibitem[{{Benson} {et~al.}(2003){Benson}, {Bower}, {Frenk}, {Lacey}, {Baugh},
  \& {Cole}}]{benson03}
{Benson}, A.~J., {Bower}, R.~G., {Frenk}, C.~S., {et~al.} 2003, \apj, 599, 38

\bibitem[{{Best} {et~al.}(2007){Best}, {von der Linden}, {Kauffmann},
  {Heckman}, \& {Kaiser}}]{best07}
{Best}, P.~N., {von der Linden}, A., {Kauffmann}, G., {Heckman}, T.~M., \&
  {Kaiser}, C.~R. 2007, \mnras, 379, 894

\bibitem[{{Binney}(1977)}]{binney77}
{Binney}, J. 1977, \apj, 215, 483

\bibitem[{{Birnboim} \& {Dekel}(2003)}]{birnboim03}
{Birnboim}, Y. \& {Dekel}, A. 2003, \mnras, 345, 349

\bibitem[{{Borisova} {et~al.}(2016){Borisova}, {Cantalupo}, {Lilly}, {Marino},
  {Gallego}, {Bacon}, {Blaizot}, {Bouch{\'e}}, {Brinchmann}, {Carollo},
  {Caruana}, {Finley}, {Herenz}, {Richard}, {Schaye}, {Straka}, {Turner},
  {Urrutia}, {Verhamme}, \& {Wisotzki}}]{borisova16}
{Borisova}, E., {Cantalupo}, S., {Lilly}, S.~J., {et~al.} 2016, \apj, 831, 39

\bibitem[{{Borthakur} {et~al.}(2013){Borthakur}, {Heckman}, {Strickland},
  {Wild}, \& {Schiminovich}}]{borthakur13}
{Borthakur}, S., {Heckman}, T., {Strickland}, D., {Wild}, V., \&
  {Schiminovich}, D. 2013, \apj, 768, 18

\bibitem[{{Bouch{\'e}} {et~al.}(2016){Bouch{\'e}}, {Finley}, {Schroetter},
  {Murphy}, {Richter}, {Bacon}, {Contini}, {Richard}, {Wendt}, {Kamann},
  {Epinat}, {Cantalupo}, {Straka}, {Schaye}, {Martin}, {P{\'e}roux},
  {Wisotzki}, {Soto}, {Lilly}, {Carollo}, {Brinchmann}, \&
  {Kollatschny}}]{bouche16}
{Bouch{\'e}}, N., {Finley}, H., {Schroetter}, I., {et~al.} 2016, \apj, 820, 121

\bibitem[{{Bouch{\'e}} {et~al.}(2006){Bouch{\'e}}, {Lehnert}, \&
  {P{\'e}roux}}]{bouche06}
{Bouch{\'e}}, N., {Lehnert}, M.~D., \& {P{\'e}roux}, C. 2006, \mnras, 367, L16

\bibitem[{{Brooks} {et~al.}(2009){Brooks}, {Governato}, {Quinn}, {Brook}, \&
  {Wadsley}}]{brooks09}
{Brooks}, A.~M., {Governato}, F., {Quinn}, T., {Brook}, C.~B., \& {Wadsley}, J.
  2009, \apj, 694, 396

\bibitem[{{Cantalupo} {et~al.}(2014){Cantalupo}, {Arrigoni-Battaia},
  {Prochaska}, {Hennawi}, \& {Madau}}]{cantalupo14}
{Cantalupo}, S., {Arrigoni-Battaia}, F., {Prochaska}, J.~X., {Hennawi}, J.~F.,
  \& {Madau}, P. 2014, \nat, 506, 63

\bibitem[{{Cen} \& {Ostriker}(2006)}]{cen06}
{Cen}, R. \& {Ostriker}, J.~P. 2006, \apj, 650, 560

\bibitem[{{Cluver} {et~al.}(2010){Cluver}, {Appleton}, {Boulanger}, {Guillard},
  {Ogle}, {Duc}, {Lu}, {Rasmussen}, {Reach}, {Smith}, {Tuffs}, {Xu}, \&
  {Yun}}]{cluver10}
{Cluver}, M.~E., {Appleton}, P.~N., {Boulanger}, F., {et~al.} 2010, \apj, 710,
  248

\bibitem[{{Codis} {et~al.}(2015){Codis}, {Pichon}, \& {Pogosyan}}]{codis15}
{Codis}, S., {Pichon}, C., \& {Pogosyan}, D. 2015, \mnras, 452, 3369

\bibitem[{{Cooper} {et~al.}(2009){Cooper}, {Bicknell}, {Sutherland}, \&
  {Bland-Hawthorn}}]{cooper09}
{Cooper}, J.~L., {Bicknell}, G.~V., {Sutherland}, R.~S., \& {Bland-Hawthorn},
  J. 2009, \apj, 703, 330

\bibitem[{{Danovich} {et~al.}(2015){Danovich}, {Dekel}, {Hahn}, {Ceverino}, \&
  {Primack}}]{danovich15}
{Danovich}, M., {Dekel}, A., {Hahn}, O., {Ceverino}, D., \& {Primack}, J. 2015,
  \mnras, 449, 2087

\bibitem[{{Dekel} \& {Birnboim}(2006)}]{dekel06}
{Dekel}, A. \& {Birnboim}, Y. 2006, \mnras, 368, 2

\bibitem[{{Dekel} {et~al.}(2013){Dekel}, {Zolotov}, {Tweed}, {Cacciato},
  {Ceverino}, \& {Primack}}]{dekel13}
{Dekel}, A., {Zolotov}, A., {Tweed}, D., {et~al.} 2013, \mnras, 435, 999

\bibitem[{{Dubois} {et~al.}(2013){Dubois}, {Pichon}, {Devriendt}, {Silk},
  {Haehnelt}, {Kimm}, \& {Slyz}}]{dubois13}
{Dubois}, Y., {Pichon}, C., {Devriendt}, J., {et~al.} 2013, \mnras, 428, 2885

\bibitem[{{Dubois} {et~al.}(2014){Dubois}, {Pichon}, {Welker}, {Le Borgne},
  {Devriendt}, {Laigle}, {Codis}, {Pogosyan}, {Arnouts}, {Benabed}, {Bertin},
  {Blaizot}, {Bouchet}, {Cardoso}, {Colombi}, {de Lapparent}, {Desjacques},
  {Gavazzi}, {Kassin}, {Kimm}, {McCracken}, {Milliard}, {Peirani}, {Prunet},
  {Rouberol}, {Silk}, {Slyz}, {Sousbie}, {Teyssier}, {Tresse}, {Treyer},
  {Vibert}, \& {Volonteri}}]{dubois14}
{Dubois}, Y., {Pichon}, C., {Welker}, C., {et~al.} 2014, \mnras, 444, 1453

\bibitem[{{Edge} {et~al.}(2010){Edge}, {Oonk}, {Mittal}, {Allen}, {Baum},
  {B{\"o}hringer}, {Bregman}, {Bremer}, {Combes}, {Crawford}, {Donahue},
  {Egami}, {Fabian}, {Ferland}, {Hamer}, {Hatch}, {Jaffe}, {Johnstone},
  {McNamara}, {O'Dea}, {Popesso}, {Quillen}, {Salom{\'e}}, {Sarazin}, {Voit},
  {Wilman}, \& {Wise}}]{edge10}
{Edge}, A.~C., {Oonk}, J.~B.~R., {Mittal}, R., {et~al.} 2010, \aap, 518, L46

\bibitem[{{Emonts} {et~al.}(2016){Emonts}, {Lehnert}, {Villar-Mart{\'{\i}}n},
  {Norris}, {Ekers}, {van Moorsel}, {Dannerbauer}, {Pentericci}, {Miley},
  {Allison}, {Sadler}, {Guillard}, {Carilli}, {Mao}, {R{\"o}ttgering}, {De
  Breuck}, {Seymour}, {Gullberg}, {Ceverino}, {Jagannathan}, {Vernet}, \&
  {Indermuehle}}]{emonts16}
{Emonts}, B.~H.~C., {Lehnert}, M.~D., {Villar-Mart{\'{\i}}n}, M., {et~al.}
  2016, Science, 354, 1128

\bibitem[{{Fall} \& {Efstathiou}(1980)}]{fall80}
{Fall}, S.~M. \& {Efstathiou}, G. 1980, \mnras, 193, 189

\bibitem[{{Ferrara} {et~al.}(2005){Ferrara}, {Scannapieco}, \&
  {Bergeron}}]{ferrara05}
{Ferrara}, A., {Scannapieco}, E., \& {Bergeron}, J. 2005, \apjl, 634, L37

\bibitem[{{Field}(1965)}]{field65}
{Field}, G.~B. 1965, \apj, 142, 531

\bibitem[{{Fragile} {et~al.}(2004){Fragile}, {Murray}, {Anninos}, \& {van
  Breugel}}]{fragile04}
{Fragile}, P.~C., {Murray}, S.~D., {Anninos}, P., \& {van Breugel}, W. 2004,
  \apj, 604, 74

\bibitem[{{Fraternali} {et~al.}(2013){Fraternali}, {Marasco}, {Marinacci}, \&
  {Binney}}]{fraternali13}
{Fraternali}, F., {Marasco}, A., {Marinacci}, F., \& {Binney}, J. 2013, \apjl,
  764, L21

\bibitem[{{Fumagalli} {et~al.}(2016){Fumagalli}, {Cantalupo}, {Dekel},
  {Morris}, {O'Meara}, {Prochaska}, \& {Theuns}}]{fumagalli16}
{Fumagalli}, M., {Cantalupo}, S., {Dekel}, A., {et~al.} 2016, \mnras, 462, 1978

\bibitem[{{Gaspari} {et~al.}(2012){Gaspari}, {Ruszkowski}, \&
  {Sharma}}]{gaspari12}
{Gaspari}, M., {Ruszkowski}, M., \& {Sharma}, P. 2012, \apj, 746, 94

\bibitem[{{Gnat} \& {Sternberg}(2007)}]{gnat07}
{Gnat}, O. \& {Sternberg}, A. 2007, \apjs, 168, 213

\bibitem[{{Goerdt} \& {Ceverino}(2015)}]{goerdt15a}
{Goerdt}, T. \& {Ceverino}, D. 2015, \mnras, 450, 3359

\bibitem[{{Gressel}(2009)}]{gressel09}
{Gressel}, O. 2009, \aap, 498, 661

\bibitem[{{Guillard} {et~al.}(2010){Guillard}, {Boulanger}, {Cluver},
  {Appleton}, {Pineau Des For{\^e}ts}, \& {Ogle}}]{guillard10}
{Guillard}, P., {Boulanger}, F., {Cluver}, M.~E., {et~al.} 2010, \aap, 518, A59

\bibitem[{{Guillard} {et~al.}(2009){Guillard}, {Boulanger}, {Pineau Des
  For{\^e}ts}, \& {Appleton}}]{guillard09}
{Guillard}, P., {Boulanger}, F., {Pineau Des For{\^e}ts}, G., \& {Appleton},
  P.~N. 2009, \aap, 502, 515

\bibitem[{{Hamer} {et~al.}(2016){Hamer}, {Edge}, {Swinbank}, {Wilman},
  {Combes}, {Salom{\'e}}, {Fabian}, {Crawford}, {Russell}, {Hlavacek-Larrondo},
  {McNamara}, \& {Bremer}}]{hamer16}
{Hamer}, S.~L., {Edge}, A.~C., {Swinbank}, A.~M., {et~al.} 2016, \mnras, 460,
  1758

\bibitem[{{Hayes} {et~al.}(2016){Hayes}, {Melinder}, {{\"O}stlin}, {Scarlata},
  {Lehnert}, \& {Mannerstr{\"o}m-Jansson}}]{hayes16}
{Hayes}, M., {Melinder}, J., {{\"O}stlin}, G., {et~al.} 2016, \apj, 828, 49

\bibitem[{{Heckman} {et~al.}(1990){Heckman}, {Armus}, \& {Miley}}]{heckman90}
{Heckman}, T.~M., {Armus}, L., \& {Miley}, G.~K. 1990, \apjs, 74, 833

\bibitem[{{Heckman} \& {Thompson}(2017)}]{heckman17}
{Heckman}, T.~M. \& {Thompson}, T.~A. 2017, ArXiv e-prints

\bibitem[{Heigl {et~al.}(2017)Heigl, Burkert, \& Gritschneder}]{Heigl2017}
Heigl, S., Burkert, A., \& Gritschneder, M. 2017, arXiv.org, arXiv:1705.03894

\bibitem[{{Hennebelle} \& {Chabrier}(2009)}]{hennebelle09}
{Hennebelle}, P. \& {Chabrier}, G. 2009, \apj, 702, 1428

\bibitem[{{Hennebelle} \& {P{\'e}rault}(1999)}]{hennebelle99}
{Hennebelle}, P. \& {P{\'e}rault}, M. 1999, \aap, 351, 309

\bibitem[{{Hopkins} {et~al.}(2012){Hopkins}, {Quataert}, \&
  {Murray}}]{hopkins12}
{Hopkins}, P.~F., {Quataert}, E., \& {Murray}, N. 2012, \mnras, 421, 3522

\bibitem[{{Hopkins} {et~al.}(2016){Hopkins}, {Torrey}, {Faucher-Gigu{\`e}re},
  {Quataert}, \& {Murray}}]{hopkins16}
{Hopkins}, P.~F., {Torrey}, P., {Faucher-Gigu{\`e}re}, C.-A., {Quataert}, E.,
  \& {Murray}, N. 2016, \mnras, 458, 816

\bibitem[{{Hu} {et~al.}(2014){Hu}, {Naab}, {Walch}, {Moster}, \& {Oser}}]{hu14}
{Hu}, C.-Y., {Naab}, T., {Walch}, S., {Moster}, B.~P., \& {Oser}, L. 2014,
  \mnras, 443, 1173

\bibitem[{{Jaffe} {et~al.}(2005){Jaffe}, {Bremer}, \& {Baker}}]{jaffe05}
{Jaffe}, W., {Bremer}, M.~N., \& {Baker}, K. 2005, \mnras, 360, 748

\bibitem[{{Kang} {et~al.}(2005){Kang}, {Ryu}, {Cen}, \& {Song}}]{kang05}
{Kang}, H., {Ryu}, D., {Cen}, R., \& {Song}, D. 2005, \apj, 620, 21

\bibitem[{{Kere{\v s}} \& {Hernquist}(2009)}]{keres09c}
{Kere{\v s}}, D. \& {Hernquist}, L. 2009, \apjl, 700, L1

\bibitem[{{Kere{\v s}} {et~al.}(2005){Kere{\v s}}, {Katz}, {Weinberg}, \&
  {Dav{\'e}}}]{keres05}
{Kere{\v s}}, D., {Katz}, N., {Weinberg}, D.~H., \& {Dav{\'e}}, R. 2005,
  \mnras, 363, 2

\bibitem[{{Kornreich} \& {Scalo}(2000)}]{kornreich00}
{Kornreich}, P. \& {Scalo}, J. 2000, \apj, 531, 366

\bibitem[{{Koyama} \& {Inutsuka}(2004)}]{koyama04}
{Koyama}, H. \& {Inutsuka}, S.-i. 2004, \apjl, 602, L25

\bibitem[{{Kritsuk} {et~al.}(2011){Kritsuk}, {Nordlund}, {Collins}, {Padoan},
  {Norman}, {Abel}, {Banerjee}, {Federrath}, {Flock}, {Lee}, {Li},
  {M{\"u}ller}, {Teyssier}, {Ustyugov}, {Vogel}, \& {Xu}}]{kritsuk11}
{Kritsuk}, A.~G., {Nordlund}, {\AA}., {Collins}, D., {et~al.} 2011, \apj, 737,
  13

\bibitem[{{Kritsuk} \& {Norman}(2002)}]{kritsuk02}
{Kritsuk}, A.~G. \& {Norman}, M.~L. 2002, \apjl, 569, L127

\bibitem[{{Laigle} {et~al.}(2015){Laigle}, {Pichon}, {Codis}, {Dubois}, {Le
  Borgne}, {Pogosyan}, {Devriendt}, {Peirani}, {Prunet}, {Rouberol}, {Slyz}, \&
  {Sousbie}}]{laigle15}
{Laigle}, C., {Pichon}, C., {Codis}, S., {et~al.} 2015, \mnras, 446, 2744

\bibitem[{{Lehnert} \& {Heckman}(1995)}]{lehnert95}
{Lehnert}, M.~D. \& {Heckman}, T.~M. 1995, \apjs, 97, 89

\bibitem[{{Lehnert} \& {Heckman}(1996)}]{lehnert96}
{Lehnert}, M.~D. \& {Heckman}, T.~M. 1996, \apj, 462, 651

\bibitem[{{Lehnert} {et~al.}(2013){Lehnert}, {Le Tiran}, {Nesvadba}, {van
  Driel}, {Boulanger}, \& {Di Matteo}}]{L13}
{Lehnert}, M.~D., {Le Tiran}, L., {Nesvadba}, N.~P.~H., {et~al.} 2013, \aap,
  555, A72

\bibitem[{{Lehnert} {et~al.}(2015){Lehnert}, {van Driel}, {Le Tiran}, {Di
  Matteo}, \& {Haywood}}]{lehnert15}
{Lehnert}, M.~D., {van Driel}, W., {Le Tiran}, L., {Di Matteo}, P., \&
  {Haywood}, M. 2015, \aap, 577, A112

\bibitem[{{Lu} {et~al.}(2015){Lu}, {Mo}, \& {Wechsler}}]{lu15}
{Lu}, Y., {Mo}, H.~J., \& {Wechsler}, R.~H. 2015, \mnras, 446, 1907

\bibitem[{{Maller} \& {Bullock}(2004)}]{maller04}
{Maller}, A.~H. \& {Bullock}, J.~S. 2004, \mnras, 355, 694

\bibitem[{{Marlowe} {et~al.}(1995){Marlowe}, {Heckman}, {Wyse}, \&
  {Schommer}}]{marlowe95}
{Marlowe}, A.~T., {Heckman}, T.~M., {Wyse}, R.~F.~G., \& {Schommer}, R. 1995,
  \apj, 438, 563

\bibitem[{{Marlowe} {et~al.}(1997){Marlowe}, {Meurer}, {Heckman}, \&
  {Schommer}}]{marlowe97}
{Marlowe}, A.~T., {Meurer}, G.~R., {Heckman}, T.~M., \& {Schommer}, R. 1997,
  \apjs, 112, 285

\bibitem[{{Martin}(1998)}]{martin98}
{Martin}, C.~L. 1998, \apj, 506, 222

\bibitem[{{Martin}(2005)}]{martin05}
{Martin}, C.~L. 2005, \apj, 621, 227

\bibitem[{{Martin} {et~al.}(2015){Martin}, {Matuszewski}, {Morrissey}, {Neill},
  {Moore}, {Cantalupo}, {Prochaska}, \& {Chang}}]{martin15}
{Martin}, D.~C., {Matuszewski}, M., {Morrissey}, P., {et~al.} 2015, \nat, 524,
  192

\bibitem[{{McCourt} {et~al.}(2012){McCourt}, {Sharma}, {Quataert}, \&
  {Parrish}}]{mcCourt12}
{McCourt}, M., {Sharma}, P., {Quataert}, E., \& {Parrish}, I.~J. 2012, \mnras,
  419, 3319

\bibitem[{{M{\'e}nard} {et~al.}(2010){M{\'e}nard}, {Scranton}, {Fukugita}, \&
  {Richards}}]{menard10}
{M{\'e}nard}, B., {Scranton}, R., {Fukugita}, M., \& {Richards}, G. 2010,
  \mnras, 405, 1025

\bibitem[{{Mo} {et~al.}(2010){Mo}, {van den Bosch}, \& {White}}]{mo10}
{Mo}, H., {van den Bosch}, F.~C., \& {White}, S. 2010, {Galaxy Formation and
  Evolution}

\bibitem[{{Navarro} {et~al.}(1997){Navarro}, {Frenk}, \& {White}}]{navarro97}
{Navarro}, J.~F., {Frenk}, C.~S., \& {White}, S.~D.~M. 1997, \apj, 490, 493

\bibitem[{{Nelson} {et~al.}(2016){Nelson}, {Genel}, {Pillepich},
  {Vogelsberger}, {Springel}, \& {Hernquist}}]{nelson16}
{Nelson}, D., {Genel}, S., {Pillepich}, A., {et~al.} 2016, \mnras, 460, 2881

\bibitem[{{Nelson} {et~al.}(2015){Nelson}, {Genel}, {Vogelsberger}, {Springel},
  {Sijacki}, {Torrey}, \& {Hernquist}}]{nelson15}
{Nelson}, D., {Genel}, S., {Vogelsberger}, M., {et~al.} 2015, \mnras, 448, 59

\bibitem[{{Nelson} {et~al.}(2013){Nelson}, {Vogelsberger}, {Genel}, {Sijacki},
  {Kere{\v s}}, {Springel}, \& {Hernquist}}]{nelson13}
{Nelson}, D., {Vogelsberger}, M., {Genel}, S., {et~al.} 2013, \mnras, 429, 3353

\bibitem[{{Ocvirk} {et~al.}(2008){Ocvirk}, {Pichon}, \& {Teyssier}}]{ocvirk08}
{Ocvirk}, P., {Pichon}, C., \& {Teyssier}, R. 2008, \mnras, 390, 1326

\bibitem[{{Ogle} {et~al.}(2010){Ogle}, {Boulanger}, {Guillard}, {Evans},
  {Antonucci}, {Appleton}, {Nesvadba}, \& {Leipski}}]{ogle10}
{Ogle}, P., {Boulanger}, F., {Guillard}, P., {et~al.} 2010, \apj, 724, 1193

\bibitem[{{Peek} {et~al.}(2015){Peek}, {M{\'e}nard}, \& {Corrales}}]{peek15}
{Peek}, J.~E.~G., {M{\'e}nard}, B., \& {Corrales}, L. 2015, \apj, 813, 7

\bibitem[{{Peterson} {et~al.}(2012){Peterson}, {Appleton}, {Helou}, {Guillard},
  {Jarrett}, {Cluver}, {Ogle}, {Struck}, \& {Boulanger}}]{peterson12}
{Peterson}, B.~W., {Appleton}, P.~N., {Helou}, G., {et~al.} 2012, \apj, 751, 11

\bibitem[{{Peterson} {et~al.}(2003){Peterson}, {Kahn}, {Paerels}, {Kaastra},
  {Tamura}, {Bleeker}, {Ferrigno}, \& {Jernigan}}]{peterson03}
{Peterson}, J.~R., {Kahn}, S.~M., {Paerels}, F.~B.~S., {et~al.} 2003, \apj,
  590, 207

\bibitem[{{Pichon} {et~al.}(2011){Pichon}, {Pogosyan}, {Kimm}, {Slyz},
  {Devriendt}, \& {Dubois}}]{pichon11}
{Pichon}, C., {Pogosyan}, D., {Kimm}, T., {et~al.} 2011, \mnras, 418, 2493

\bibitem[{{Pinto} {et~al.}(2014){Pinto}, {Fabian}, {Werner}, {Kosec},
  {Ahoranta}, {de Plaa}, {Kaastra}, {Sanders}, {Zhang}, \&
  {Finoguenov}}]{pinto14}
{Pinto}, C., {Fabian}, A.~C., {Werner}, N., {et~al.} 2014, \aap, 572, L8

\bibitem[{{Price}(2012)}]{price12}
{Price}, D.~J. 2012, \mnras, 420, L33

\bibitem[{{Rafferty} {et~al.}(2008){Rafferty}, {McNamara}, \&
  {Nulsen}}]{rafferty08}
{Rafferty}, D.~A., {McNamara}, B.~R., \& {Nulsen}, P.~E.~J. 2008, \apj, 687,
  899

\bibitem[{{Raymond}(1979)}]{raymond79}
{Raymond}, J.~C. 1979, \apjs, 39, 1

\bibitem[{{Salom{\'e}} {et~al.}(2011){Salom{\'e}}, {Combes}, {Revaz}, {Downes},
  {Edge}, \& {Fabian}}]{salome11}
{Salom{\'e}}, P., {Combes}, F., {Revaz}, Y., {et~al.} 2011, \aap, 531, A85

\bibitem[{{Sharma} {et~al.}(2012{\natexlab{a}}){Sharma}, {McCourt}, {Parrish},
  \& {Quataert}}]{sharma12a}
{Sharma}, P., {McCourt}, M., {Parrish}, I.~J., \& {Quataert}, E.
  2012{\natexlab{a}}, \mnras, 427, 1219

\bibitem[{{Sharma} {et~al.}(2012{\natexlab{b}}){Sharma}, {McCourt}, {Quataert},
  \& {Parrish}}]{sharma12}
{Sharma}, P., {McCourt}, M., {Quataert}, E., \& {Parrish}, I.~J.
  2012{\natexlab{b}}, \mnras, 420, 3174

\bibitem[{{Sharma} {et~al.}(2010){Sharma}, {Parrish}, \& {Quataert}}]{sharma10}
{Sharma}, P., {Parrish}, I.~J., \& {Quataert}, E. 2010, \apj, 720, 652

\bibitem[{{Singh} \& {Sharma}(2015)}]{singh15}
{Singh}, A. \& {Sharma}, P. 2015, \mnras, 446, 1895

\bibitem[{{Suresh} {et~al.}(2015){Suresh}, {Bird}, {Vogelsberger}, {Genel},
  {Torrey}, {Sijacki}, {Springel}, \& {Hernquist}}]{suresh15}
{Suresh}, J., {Bird}, S., {Vogelsberger}, M., {et~al.} 2015, \mnras, 448, 895

\bibitem[{{Sutherland} {et~al.}(2003){Sutherland}, {Bicknell}, \&
  {Dopita}}]{sutherland03}
{Sutherland}, R.~S., {Bicknell}, G.~V., \& {Dopita}, M.~A. 2003, \apj, 591, 238

\bibitem[{{Sutherland} \& {Dopita}(1993)}]{sutherland93}
{Sutherland}, R.~S. \& {Dopita}, M.~A. 1993, \apjs, 88, 253

\bibitem[{{Thompson} {et~al.}(2016){Thompson}, {Quataert}, {Zhang}, \&
  {Weinberg}}]{thompson16}
{Thompson}, T.~A., {Quataert}, E., {Zhang}, D., \& {Weinberg}, D.~H. 2016,
  \mnras, 455, 1830

\bibitem[{{Tillson} {et~al.}(2015){Tillson}, {Devriendt}, {Slyz}, {Miller}, \&
  {Pichon}}]{tillson15}
{Tillson}, H., {Devriendt}, J., {Slyz}, A., {Miller}, L., \& {Pichon}, C. 2015,
  \mnras, 449, 4363

\bibitem[{{Tremblay} {et~al.}(2012){Tremblay}, {O'Dea}, {Baum}, {Clarke},
  {Sarazin}, {Bregman}, {Combes}, {Donahue}, {Edge}, {Fabian}, {Ferland},
  {McNamara}, {Mittal}, {Oonk}, {Quillen}, {Russell}, {Sanders}, {Salom{\'e}},
  {Voit}, {Wilman}, \& {Wise}}]{tremblay12}
{Tremblay}, G.~R., {O'Dea}, C.~P., {Baum}, S.~A., {et~al.} 2012, \mnras, 424,
  1026

\bibitem[{{Tumlinson} {et~al.}(2011){Tumlinson}, {Thom}, {Werk}, {Prochaska},
  {Tripp}, {Weinberg}, {Peeples}, {O'Meara}, {Oppenheimer}, {Meiring}, {Katz},
  {Dav{\'e}}, {Ford}, \& {Sembach}}]{tumlinson11}
{Tumlinson}, J., {Thom}, C., {Werk}, J.~K., {et~al.} 2011, Science, 334, 948

\bibitem[{{van de Voort} \& {Schaye}(2012)}]{vandeVoort12}
{van de Voort}, F. \& {Schaye}, J. 2012, \mnras, 423, 2991

\bibitem[{{van de Voort} {et~al.}(2011){van de Voort}, {Schaye}, {Booth}, \&
  {Dalla Vecchia}}]{vandeVoort11}
{van de Voort}, F., {Schaye}, J., {Booth}, C.~M., \& {Dalla Vecchia}, C. 2011,
  \mnras, 415, 2782

\bibitem[{{Vernet} {et~al.}(2017){Vernet}, {Lehnert}, {De Breuck},
  {Villar-Mart{\'{\i}}n}, {Wylezalek}, {Falkendal}, {Drouart}, {Kolwa},
  {Humphrey}, {Venemans}, \& {Boulanger}}]{vernet17}
{Vernet}, J., {Lehnert}, M.~D., {De Breuck}, C., {et~al.} 2017, \aap, 602, L6

\bibitem[{{Voit} {et~al.}(2015{\natexlab{a}}){Voit}, {Bryan}, {O'Shea}, \&
  {Donahue}}]{voit15}
{Voit}, G.~M., {Bryan}, G.~L., {O'Shea}, B.~W., \& {Donahue}, M.
  2015{\natexlab{a}}, \apjl, 808, L30

\bibitem[{{Voit} {et~al.}(2015{\natexlab{b}}){Voit}, {Donahue}, {Bryan}, \&
  {McDonald}}]{voit15a}
{Voit}, G.~M., {Donahue}, M., {Bryan}, G.~L., \& {McDonald}, M.
  2015{\natexlab{b}}, \nat, 519, 203

\bibitem[{{Welker} {et~al.}(2014){Welker}, {Devriendt}, {Dubois}, {Pichon}, \&
  {Peirani}}]{welker14}
{Welker}, C., {Devriendt}, J., {Dubois}, Y., {Pichon}, C., \& {Peirani}, S.
  2014, \mnras, 445, L46

\bibitem[{{Werk} {et~al.}(2014){Werk}, {Prochaska}, {Tumlinson}, {Peeples},
  {Tripp}, {Fox}, {Lehner}, {Thom}, {O'Meara}, {Ford}, {Bordoloi}, {Katz},
  {Tejos}, {Oppenheimer}, {Dav{\'e}}, \& {Weinberg}}]{werk14}
{Werk}, J.~K., {Prochaska}, J.~X., {Tumlinson}, J., {et~al.} 2014, \apj, 792, 8

\bibitem[{{Wetzel} \& {Nagai}(2015)}]{wetzel15}
{Wetzel}, A.~R. \& {Nagai}, D. 2015, \apj, 808, 40

\bibitem[{{White} \& {Rees}(1978)}]{white78}
{White}, S.~D.~M. \& {Rees}, M.~J. 1978, \mnras, 183, 341

\bibitem[{{Zhuravleva} {et~al.}(2014){Zhuravleva}, {Churazov}, {Schekochihin},
  {Allen}, {Ar{\'e}valo}, {Fabian}, {Forman}, {Sanders}, {Simionescu},
  {Sunyaev}, {Vikhlinin}, \& {Werner}}]{zhuravleva14}
{Zhuravleva}, I., {Churazov}, E., {Schekochihin}, A.~A., {et~al.} 2014, \nat,
  515, 85

\end{thebibliography}

\begin{appendix}
\section{Parameters in the model}

The quantities that are important in setting the initial conditions
of the stream-ambient halo gas interaction are the mass, virial
velocity, and virial radius of the dark matter halo which we denote
as $M_\mathrm{H}$, $v_\mathrm{vir}$, and $r_\mathrm{vir}$. The dark
matter distribution is given by a NFW profile with a concentration
parameter, c, of 10 \citep{navarro97}. The halo is filled by a hot gas
of temperature $T_\mathrm{H}$, which we assume to be equal to the virial
temperature of the halo, $T_\mathrm{vir}$. The density of the hot halo,
$\rho_\mathrm{H}$, is assumed to follow that of the dark matter density
with radius, but is multiplied by the cosmological baryon density relative
to the dark matter density, $f_\mathrm{B}=0.18$.  This is $\approx
37 f_\mathrm{B}\rho_\mathrm{crit}$, where $\rho_\mathrm{crit}$ is the
critical density of the Universe.  The halo pressure, $P_\mathrm{H}$,
is related to $T_\mathrm{H}$ and $\rho_\mathrm{H}$. The filling factor
of this gas is assumed to be one. We are agnostic about how this hot,
high volume-filling factor halo at the virial temperature formed but
note that it likely forms by a combination of accretion of gas from the
intergalactic medium and heating through the radiative and mechanical energy
output of the galaxy embedded in the halo \citep[e.g.,][]{suresh15, lu15}.

\begin{table}[ht]
\caption{Halo and gas parameters for example in \S\,\ref{sec:specificcase}}
\label{tab:haloparams}
\begin{tabular}{l l c}
\hline
\hline
Parameter Name & Symbol & Value \\
\hline
Halo mass & $M_\mathrm{H}$&  $10^{13}\,\mathrm{M}_{\sun}$\\
Baryonic fraction & $f_\mathrm{B}$ & 0.18\\
Redshift & z &  2 \\
Virial radius & $r_\mathrm{vir}$&  220 kpc\\
Virial velocity & $v_\mathrm{vir}$&  440 $\mathrm{km\,s^{-1}}$\\
Critical density & $\rho_\mathrm{crit}$/$\mu m_\mathrm{p}$& 7.6$\times$10$^{-5}$ cm$^{-3}$\\
Number density at $r_\mathrm{vir}$ & $n_\mathrm{0}$& 5.1$\times$10$^{-4}$ cm$^{-3}$\\
\hline
Adiabatic index & $\gamma$ & $5/3$\\
Mean particle mass & $\mu m_\mathrm{p}$ & $0.6\times m_\mathrm{p}$\\
\hline
\end{tabular}
\end{table}

\begin{table*}
\caption{Model variables and their relationships\tablefootmark{$\dagger$}.}
\label{tab:modelvars}
\begin{tabular}{l r l c}
\hline
\hline
Variable Name & Symbol & Equation\\
\hline
Temperature floor & $T_0$ & $=10^4\,\mathrm{K}$\\
Initial speed of sound & $c_1$ & $=\sqrt{\gamma k_\mathrm{B}T_0/\mu m_\mathrm{p}}$\\
Incoming Mach number & $\mathcal{M}_1$ & $=v_\mathrm{vir}/c_1$ \\
\hline
Density of the halo gas at $r_\mathrm{vir}$ & $\rho_\mathrm{H}$ & $=\rho_\mathrm{NFW}\left(r_\mathrm{vir}\right)\approx 37.0\times f_\mathrm{B}\rho_\mathrm{crit}\left(z\right)$ &  \\
Temperature of the halo gas & $T_\mathrm{H}$ & $=T_\mathrm{vir}=\mu m_\mathrm{p} v_\mathrm{vir}^2(\gamma-1)/2k_\mathrm{B}$\\
Pressure of the halo gas at $r_\mathrm{vir}$ & $P_\mathrm{H}$ & $=k_\mathrm{B}T_\mathrm{H}\rho_\mathrm{H}/\mu m_\mathrm{p}$\\
\hline
Density of the post-sock gas\tablefootmark{a} & $\rho_\mathrm{ps}$ & $=\left(\gamma+1\right)/\left(\gamma-1\right)\mathcal{M}_1^2/\left[\mathcal{M}_1^2+2/(\gamma-1)\right]\rho_0$\\
Pressure of the post-shock gas\tablefootmark{a} & $P_\mathrm{ps}$ & $=\left(\gamma-1\right)/\left(\gamma+1\right)\left[2\gamma/(\gamma-1)\mathcal{M}_1^2-1\right] P_0$\\
Post-shock speed of sound & $c_\mathrm{ps}$ & $=\gamma P_\mathrm{ps}/\rho_\mathrm{ps}$\\
\hline
Temperature of the warm phase & $T_\mathrm{w}$ & $=T_0$\\
Density of the warm phase & $\rho_\mathrm{w}$ & $=\rho_\mathrm{H}T_\mathrm{H}/T_\mathrm{w}$\\
\hline
Density of the hot phase (post-expansion) & $\rho_\mathrm{h}$ & $=\rho_\mathrm{ps}\left(P_\mathrm{ps}/P_\mathrm{H}\right)^{1/\gamma}$\\
Temperature of the hot phase (post-expansion) & $T_\mathrm{h}$ & $=\rho_\mathrm{H}T_\mathrm{H}/\rho_\mathrm{h}$\\
\hline
Volume-averaged density of the warm phase & $\left\langle\rho_\mathrm{w}\right\rangle_\mathrm{v}$ & $=\phi_\mathrm{v,w}\rho_\mathrm{w}$\\
Volume-averaged density & $\rho_2$ & $=\phi_\mathrm{v,w}\rho_\mathrm{w}+(1-\phi_\mathrm{v,w}\rho_\mathrm{h})$\\
Expansion factor of the post-shock stream\tablefootmark{b} & $S$ & $=\left(1-\sqrt{\Delta}\right)/\left[(\gamma-1)/f+2\eta)\right]$\\
Velocity dispersion of the warm clouds & $\sigma_\mathrm{turb}$ & $=\sqrt{2k_\mathrm{B}T_\mathrm{w}\eta f/(\gamma-1)\phi_\mathrm{v,w}\mu m_\mathrm{p}}$\\
\hline
Halo dynamical time & $t_\mathrm{dyn,halo}$ & $=r_\mathrm{vir}/v_\mathrm{vir}$\\
Cooling time of the phase $\Phi\in\left\{\mathrm{ps,w,h}\right\}$ & $t_\mathrm{cool,\Phi}$ & $=k_\mathrm{B}\mu m_\mathrm{p}T_\Phi/\rho_\Phi\Lambda\left(T_\Phi\right)$\\
Expansion time of the post-shock gas& $t_\mathrm{expand}$ & $=2\gamma r_\mathrm{stream}/c_\mathrm{ps}$\\
Disruption time of the turbulent warm phase & $t_\mathrm{disrupt}$ & $=r_\mathrm{stream}/\sigma_\mathrm{turb}$\\
\hline
Isobaric cooling length $\Phi\in\left\{\mathrm{ps,w,h}\right\}$ & $\lambda_\mathrm{cooling}$ & $=c_\mathrm{\Phi}t_\mathrm{cool,\Phi}$\\
\hline
\end{tabular}
\tablefoot{\\
\tablefoottext{$\dagger$}{A graphical representation of many of these variables is shown in Fig.~\ref{fig:sketch}.}\\
\tablefoottext{a}{Standard normal shock equation from the Rankine-Hugoniot jump conditions.}\\
\tablefoottext{b}{The equation assumes $\Delta=1-4\left[\eta f+(\gamma-1)/2\right]\rho_\mathrm{H}/\rho_2$}
}
\end{table*}

The gas accretes through a stream of infalling gas with radius,
$r_\mathrm{stream}$, we assume that it passes through a shock and that
the properties of the post-shock gas is given by the standard set of
shock equations.  We simply scale the density of the accreting stream
by a factor, $f$, which is its density contrast of the background dark
matter density at the virial radius multiplied by the cosmological
baryon fraction (i.e., $\rho_\mathrm{H}$).  We further assume that there
is a temperature floor in the post-shock gas of $10^4\,\mathrm{K}$.
We assumed this temperature mainly because we also assume that the
metallicity of the accreting stream is 10$^{-3}$ of the solar value.
The gas cannot cool much beyond $10^4\,\mathrm{K}$ due to it lacking
heavy metals (and is likely heated by the meta-galactic flux and the
ionizing field of the galaxy embedded in the halo).  This assumption,
although naive, is also extremely conservative in that this implies
the post-shock gas will have one of the longest possible radiative
cooling time \citep[see][]{sutherland93, gnat07}.  We use the cooling
curve, $\Lambda(T)$, from \citet{gnat07} to compute the cooling times
in the post-shock gas. We assume that the temperature of the gas in
the stream before passing through the shock is also $10^4\,\mathrm{K}$
($T_1$).  At those temperatures and very low metallicity, we assume that
no molecules form, so that the adiabatic index of the gases is always
that of a monatomic gas, namely $\gamma=5/3$.

The parameters we use in the model, given our assumed mass and
redshift are given in Table~\ref{tab:haloparams}.  We enumerate
for completeness and clarity all variables used in our analysis in
Table~\ref{tab:modelvars}.

\end{appendix}

\end{document}